\documentclass[aps,pre,floatfix,twocolumn]{revtex4-1}
\usepackage[dvips]{graphicx}
\usepackage{amsmath}
\usepackage{verbatim} 
\usepackage{color}
\usepackage{balance} 
\newcommand{\etalia}{{\it et al.~}}
\newcommand{\la}{\left\langle}
\newcommand{\ra}{\right\rangle}

\begin{document}

\title{Response of Hydrogels and Microgels to Nanoparticle Crowding}

\author{Mahesh Aryal and Alan R.~Denton}
\email[]{alan.denton@ndsu.edu}
\affiliation{Department of Physics, North Dakota State University, Fargo, ND 58108-6050, USA}

\begin{abstract}
Internal degrees of freedom of cross-linked polymer networks enable compressible hydrogels and
microgels to swell or deswell by absorbing or expelling solvent in response to environmental changes. 
Macromolecules in solution can influence the swelling response by acting as crowding agents, with 
practical relevance for drug delivery and biosensing. We extend the Flory-Rehner theory of polymer 
network swelling to incorporate nanoparticle crowders as another implicit species that can perturb 
a gel through volume exclusion, including the entropic cost of crowder penetration from polymer 
field theory. Within the free energy landscape of asymmetric ternary mixtures, we numerically 
solve equilibrium relations equating solvent and crowder chemical potentials inside and outside 
the gel. Independently, we perform Monte Carlo simulations, including trial moves that allow gels to 
swell and take up crowders. We investigate how crowder size, solution concentration, cross-link density,
and solvent quality influence swelling of polymeric gels, partitioning of crowders between 
inside and outside, and structural properties of microgel suspensions governed by Hertzian elastic pair 
interactions. Our results show that nanoparticle crowding promotes deswelling, thereby weakening 
interparticle correlations, and that, with increasing the concentration of crowders, polymeric gels
reach a maximum loading capacity, dependent on crowder size and polymer cross-link density.
\end{abstract}
\maketitle


\section{Introduction}\label{intro}
Hydrogels are cross-linked polymer networks, whose internal degrees of freedom enable them 
to swell to many times their dry size by absorbing solvent (e.g., water)~\cite{HydrogelBook2012}.
Microgels are soft colloidal hydrogel particles, typically tens of nanometers to a micron in size, 
whose translational degrees of freedom allow them to self-assemble into ordered structures
\cite{MicrogelBook2011,schurtenberger-ZPC2012,karg-richtering2019,scotti2022}.
Such polymeric gels swell and deswell in response to changes in solution properties, 
such as temperature, pH, and solvent composition. 
Their inherent capacity to undergo significant volume changes in response to external stimuli
qualifies microgels as smart colloidal particles and enables a wide range of applications, 
from targeted drug delivery systems to advanced biosensors, where precise control over material 
properties can enhance functionality and performance~\cite{malmsten2011,plamper2017functional,
scheffold2020pathways}.

Changing solvent composition by adding cosolvents or components of higher molecular weight, 
such as macromolecules or nanoparticles, can influence polymer conformations and swelling behavior.
Mixing cosolvents in solution, for instance, can alter solvent quality, thus modifying effective 
interactions between polymer segments and the statistical distribution of segments, e.g., changing 
a polymer coil from a random walk to a self-avoiding walk. Adding nanoscale particles can crowd 
other species through excluded-volume interactions~\cite{schaefer1999}.

The phenomenon of macromolecular crowding manifests when the conformations of biopolymers 
(e.g., proteins, RNA, DNA) or other macromolecules in biological cells are constrained by 
interactions with other species or by confinement~\cite{minton1981,minton1992,minton1997,minton2000,
minton2005,ellis2001,cheung2005,phillips2009,elcock2010,hancock2012,denton-cmb2013}. 
Similarly, the conformations of polymer chains in a swollen gel can collectively respond 
to the presence of crowding agents (crowders) in solution. Importantly, since gels are penetrable, 
the crowding here comes from within. The main purpose of this paper is to model and analyze 
the swelling response of hydrogels and microgels to macromolecular crowding and, conversely, the 
influence of swelling on uptake and release of crowders, which could represent drug or dye molecules.

The swelling behavior of polymer networks in pure solvents has been traditionally modeled using 
the classical Flory-Huggins and Flory-Rehner theories~\cite{Flory-Rehner1943-I,Flory-Rehner1943-II,
flory1953principles,doi2013soft}, which attribute swelling to a balance between osmotic pressure of 
a binary polymer-solvent mixture and elastic pressure originating from the entropic cost of deforming 
polymer chains. Previous studies have extended the Flory-Rehner theory to ternary mixtures and have 
modeled polydimethylsiloxane (PDMS)~\cite{favre1995multicomponent}, cross-linked polymer membranes
~\cite{okeowo2006multicomponent}, copolymer networks in mixed solvents~\cite{godbole2017swelling}, 
and polydisperse mixtures of polymers~\cite{stone2024}, as well as microgels mixed with 
free polymers~\cite{routh2006}. 

Several recent experimental studies have measured properties of microgels immersed in macromolecular
solutions. For example, Bradley \etalia~\cite{bradley-vincent2005,routh2006} investigated swelling and 
uptake properties of copolymer microgels of N-isopropylacrylamide (NIPAM) and acrylic acid in aqueous 
solutions of poly(ethylene oxide) (PEO) of varying molecular weights.
In a relevant drug delivery application, lipid nanoparticles were encapsulated in dextran microgels 
and used as a chemotherapeutic treatment against colon cancer~\cite{chiu2022}.
In another study, nanogels were used as sacrificial pH-degradable templates to control 
the porosity of functional dextran-based microgels~\cite{pich2023}.

Despite progress, a comprehensive theoretical framework for modeling the interplay between network 
swelling and the loading and expulsion of a third volume-excluding component is lacking. Our research 
aims to bridge this gap by extending the Flory-Rehner theory to mixtures of polymer, solvent, and 
substantially larger particles (crowders) that significantly exclude volume, with the goal of 
providing deeper insight into the swelling behavior and functional capabilities of polymeric gels.

As a practical implementation of our theory, we also generalize a Monte Carlo simulation approach 
previously developed for modeling suspensions of interacting microgels in pure solvents 
\cite{urich2016swelling,weyer2018concentration,brito2019modeling}.
The simulations include novel trial moves that allow microgels to change both size and composition, 
i.e., interior concentration of crowders. In this way, we investigate how the swelling behavior 
of microgels depends on cross-link density, solvent quality, and crowder size and concentration.

The remainder of the paper is organized as follows. In Sec.~\ref{models}, we describe our model of soft,
compressible hydrogels and colloidal microgels dispersed in a solution of hard, nanoscale crowding agents.
We then derive our theory of polymeric gel swelling and absorption (uptake) of crowders from solution
by extending the Flory-Rehner theory of cross-linked polymer networks to predict the free energy of 
ternary polymer-solvent-crowder mixtures.
A distinguishing feature of our model and theory is the incorporation of excluded volume of the crowders.
From the approximate free energy, we derive expressions for the chemical potentials of the 
solvent and crowders. In Sec.~\ref{methods}, we describe our computational methods for equating the 
chemical potentials of the respective species inside and outside the gel. By numerically solving two 
nonlinear equations, we obtain the equilibrium swelling ratio and interior crowder volume fraction. 
We also develop and perform Monte Carlo simulations of one or multiple microgels, including elastic 
interparticle interactions, and compute single-microgel equilibrium properties, as well as structural 
properties of a suspension of microgels. In Sec.~\ref{results}, we present and compare our numerical 
results and simulation data and discuss our interpretation of the impact of crowders on polymeric gels, 
and vice versa, of gels on the distribution of crowders. Finally, we summarize and conclude.

\section{Models and Theories}\label{models}
\subsection{Model of Polymeric Gel-Crowder Solutions}
\begin{figure}
\includegraphics[width=0.9\columnwidth,angle=0]{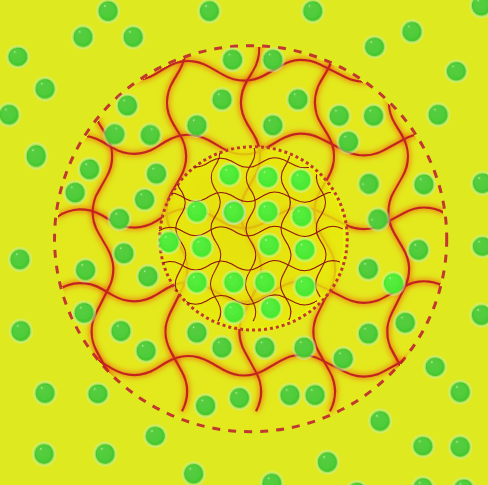}
\vspace*{-0.2cm}
\caption{A swollen microgel, consisting of a cross-linked polymer network (red chains, bounded by 
outer dashed circle), immersed in a crowder solution, deswells to a collapsed state 
(inner dotted circle) by expelling solvent (yellow background) and crowders (green particles).
Created with BioRender.com/1kz4x5k.}
\label{SingleMicrogelSwelling}
\end{figure}

\begin{figure}
\includegraphics[width=0.9\columnwidth,angle=0]{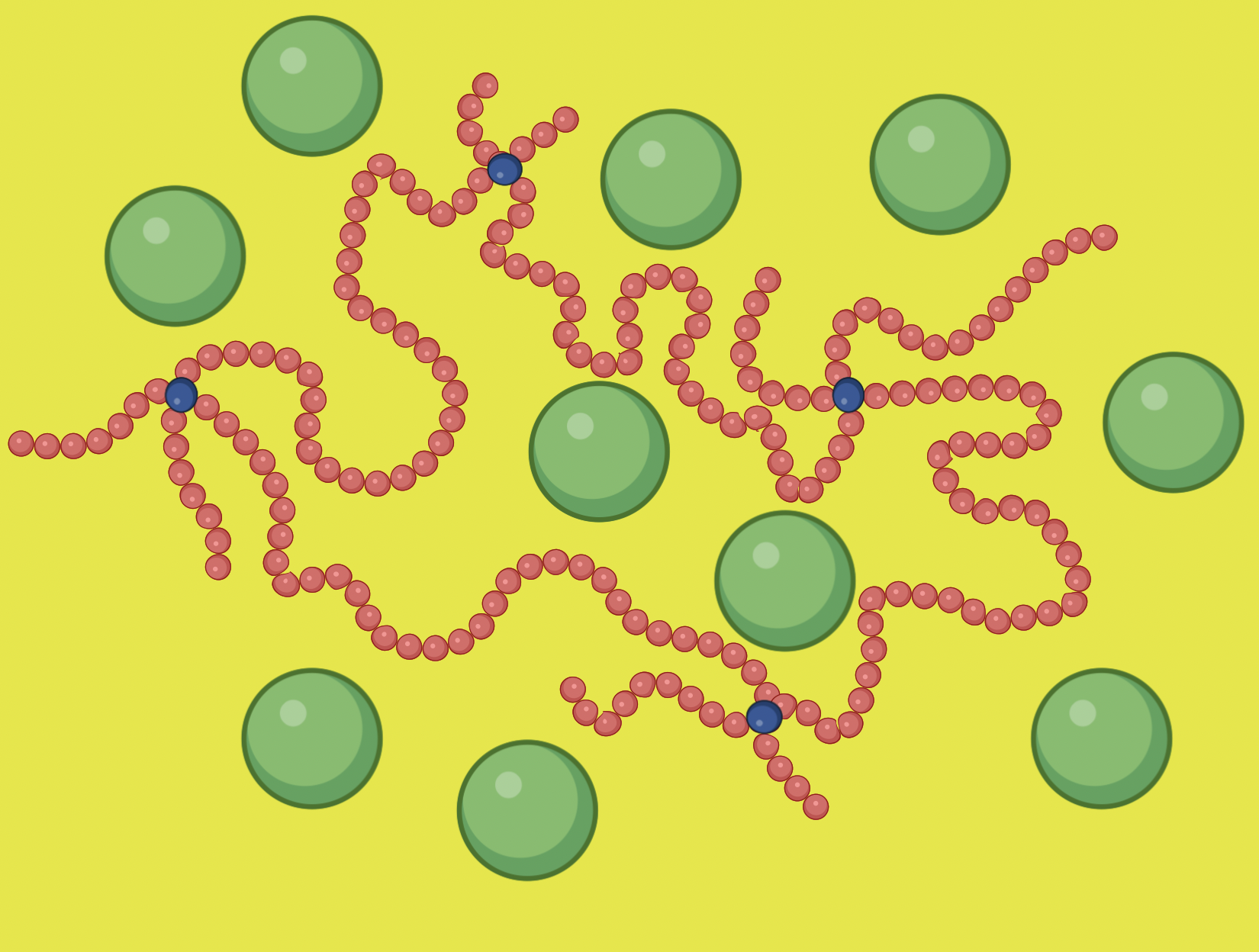}
\vspace*{-0.2cm}
\caption{Microscopic depiction of the interior of a microgel, consisting of a network of 
polymer chains (red), connected by cross-linkers (blue spheres), immersed in a 
solution of solvent (yellow background) and crowders (green spheres).
Created with BioRender.com/b80l719.}
\label{MeshNanoparticle}
\end{figure}

The system of interest consists of a polymeric gel and crowding agents (crowders) dispersed in a solvent 
at constant volume $V$ and temperature $T$, as depicted in Fig.~\ref{SingleMicrogelSwelling}.
The term ``crowder'' is used generically here to describe any nanoscale object that excludes volume to the 
microgels, e.g., small proteins, nanoparticles, or other macromolecules smaller than the pores of the gel. 
We assume a suspension of $N_m$ microgels can exchange solvent and crowders via a semi-permeable membrane
with a reservoir of crowder solution. The number of solvent molecules $N_s$ and of crowders $N_c$ vary 
through exchange with the reservoir, which imposes chemical potentials of those species. The microgels 
themselves are assumed permeable to solvent, but only selectively permeable to crowders, limited by an 
entropic cost of perturbing the polymer network (see Fig.~\ref{MeshNanoparticle}).
 
\begin{figure}
\includegraphics[width=0.9\columnwidth,angle=0]{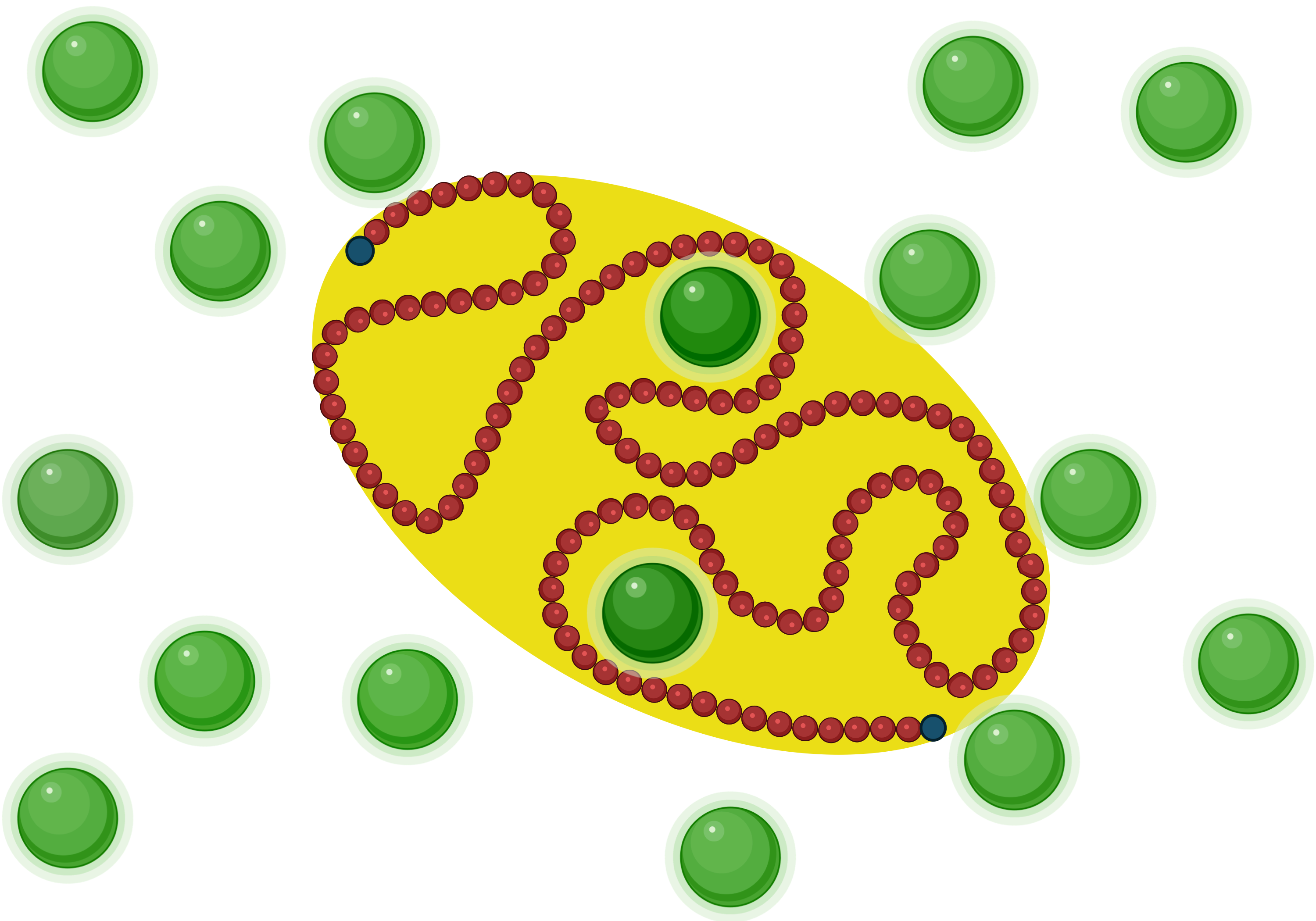}
\vspace*{-0.2cm}
\caption{Spanning cross-links (blue spheres), the segments of a polymer chain (red spheres) form a 
random walk, whose shape can be approximated by an ellipsoid. Assuming affine swelling, the 
polymer chain expands or contracts by a factor proportional to the network’s swelling ratio, 
which is influenced by the presence of crowders (green spheres).
Created with BioRender.com/4ywc5yl.}
\label{polyinnano}
\end{figure}

We model microgels as spherical, compressible, soft colloidal particles, each composed of a cross-linked 
network of polymer chains, consisting of $N_\text{mon}$ monomers, $N_\text{ch}$ chains, and tetravalent 
cross-links with fraction $x=0.5N_\text{ch}/N_\text{mon}$. In experiments, microgels typically have 
a core-shell structure, consisting of a densely cross-linked core surrounded by a more loosely 
cross-linked shell~\cite{zaccarelli2024,moncho-jorda2025}. In the present work, however, we adopt
a simplified description with a uniform distribution of cross-links. Such homogeneous microgels 
can be synthesized under appropriate conditions~\cite{pelton2011,varga2011,karg-richtering2019}.
Polymer chains comprise statistically independent segments (each a few monomers) of length $l$ 
(Kuhn length), a measure of the chain stiffness~\cite{doi2013soft,deGennes1979}. In a good solvent, 
the segments avoid one another, but otherwise rotate freely~\cite{doi2013soft,rubinstein2003polymer}.
This flexibility allows a chain to adopt a vast number of conformations, whose statistics closely 
follow a self-avoiding random walk (SAW) and whose shapes can be modeled as fluctuating, penetrable 
ellipsoids~\cite{kuhn1934shape,rudnick1987shapes,murat1998many,eurich2001soft,lim-denton-JCP2014,
lim-denton-JCP2016,lim-denton-SM2016,davis-dentonJCP2018} (see Fig.~\ref{polyinnano}).
An uncrowded SAW chain with $N_\text{seg}=N_\text{mon}/N_\text{ch}$ segments, spanning neighboring 
cross-links, has a root-mean-square radius of gyration~\cite{sciutto1994,sciutto1996,davis-dentonJCP2018}
\begin{equation}
R_g = C N_\text{seg}^{\gamma}l,
\label{Rg}
\end{equation}
with Flory exponent $\gamma=0.588$ and amplitude $C=0.441$.
Internal degrees of freedom allow a microgel to absorb/expel surrounding solution and swell/deswell 
between a swollen state, with radius $a$ and volume $V_m$, and collapsed (dry or unswollen) state, 
with radius $a_0$ and volume $V_0$. The degree of swelling is characterized by the linear swelling ratio,
$\alpha=a/a_0=(V_m/V_0)^{1/3}=\phi_p^{-1/3}$, where $\phi_p=V_0/V_m$ defines the fraction of the
total volume occupied by polymer monomers.

To isolate the influence of excluded volume on microgel swelling, we model crowders
as mutually impenetrable hard spheres of radius $R_c$ interacting via a pair potential
\begin{equation}
v_\text{HS}(r) =
\begin{cases}
\infty, & r < 2R_c
\\[2ex]
0, & r\ge 2R_c.
\end{cases}  
\label{vhs}
\end{equation}
Assuming polymer monomers and solvent molecules to be much smaller than the crowders, 
the latter thus exclude to the former a volume $v_c=(4\pi/3)R_c^3$ and to other crowders
a volume $8v_c$. As we shall see, excluded volume affects swelling in two independent ways 
-- via mixing entropy and polymer conformational entropy.

\subsection{Theory of Polymeric Gel-Crowder Solutions}
To model polymeric gels suspended in a solution of crowders, we adapt the Flory-Rehner theory 
of polymer networks~\cite{Flory-Rehner1943-I,Flory-Rehner1943-II,flory1953principles}
to approximate the free energy of a ternary mixture of polymer, solvent, and crowders. 
Broad applications to swelling of polymeric gels have demonstrated the utility and qualitative
accuracy of the Flory-Rehner theory, as well as some limitations~\cite{richtering2017}.
We initially focus on a bulk hydrogel or single isolated microgel and then generalize to a suspension 
of interacting microgels. Starting from the Flory-Huggins theory of polymer solutions~\cite{doi2013soft}, 
we consider a lattice model of a ternary mixture, in which the lattice sites may be singly occupied 
by polymer monomers or solvent molecules or multiply occupied by crowders. To account for 
the potentially much larger volume of crowders, relative to polymer monomers and solvent molecules,
we allow the crowders to multiply occupy a number of sites $n=v_c/v_p>1$, where $v_c$, $v_p$, 
and $v_s$ are respective volumes of crowders, polymer monomers, and solvent molecules.

The free energy of a polymeric gel in a solution of crowders is a combination of mixing entropy,
associated with freedom of the different species of particle to assemble on the lattice, 
elastic entropy, associated with swelling of the polymer network, pair interaction energy of mixing 
of the various species, and a free energy of penetration of the polymer network by crowders. 
The mixing entropy $S_\text{mix}$ depends, through the Boltzmann relation, 
$S_\text{mix}=k_B\ln\Omega_\text{mix}$, on the total number of configurations $\Omega_\text{mix}$ 
of all the particles ($k_B$ is the Boltzmann constant). For a sufficiently dilute solution, 
the placement of crowders can be decoupled from the placement of the other species, allowing 
$\Omega_\text{mix}$ to be factorized, according to $\Omega_\text{mix}=\Omega_c\Omega_{ps}$, 
into the respective numbers of ways of placing $N_c$ crowders, $N_p$ polymer monomers, and 
$N_s$ solvent molecules. Assuming full occupancy of the lattice, these numbers are not all 
independent, but subject to the constraint
\begin{equation}
N_p+N_s+n N_c=M,
\label{M}
\end{equation}
where $M$ is the total number of lattice sites, which changes as the gel swells or deswells.
Taking the polymer monomer volume as a unit of volume, and expressing the gel volume as 
\begin{equation}
V_m=Mv_p=N_pv_p+N_sv_p+N_cv_c,
\label{Vm}
\end{equation}
the site occupation (volume) fractions of the various species are related via
\begin{equation}
\phi_p+\tilde\phi_s+\phi_c=1,
\label{phisum}
\end{equation}
where $\phi_p=N_p/M=N_pv_p/V_m$, $\phi_s=N_sv_s/V_m$, and $\phi_c=N_cv_c/V_m$ are the polymer, solvent, 
and crowder volume fractions, respectively, and for convenience we define 
$\tilde\phi_s\equiv N_s/M=(v_p/v_s)\phi_s$ as an effective solvent volume fraction that accounts for 
different polymer and solvent volumes. In the reservoir of crowder solution (absent polymer), 
Eq.~(\ref{phisum}) reduces to
\begin{equation}
\tilde\phi_s^0+\phi_c^0=1,
\label{phisum0}
\end{equation}
where $\tilde\phi_s^0$ and $\phi_c^0$ are the respective volume fractions of solvent and
crowders in the reservoir.

To incorporate mutual excluded volume, we assume that each crowder excludes to solvent and polymer 
$n$ sites, but to other crowders a number of sites $n_\text{ex}>n$. 
For a solution sufficiently dilute that overlaps of crowder excluded-volume spheres may be neglected 
in the counting of microstates, the number of ways of placing $N_c$ indistinguishable crowders 
is approximated as
\begin{equation}
\Omega_c^0 = \frac{M}{N_c!}(M-n_\text{ex})(M-2n_\text{ex})\cdots[M-(N_c-1)n_\text{ex}].
\label{Omegac0}
\end{equation}
Assuming $N_c\gg 1$ and applying the Stirling approximation, the associated crowder entropy is given by
\begin{eqnarray}
\frac{S_c^0}{k_B} &=& -N_c\ln\left(\frac{N_c}{M}\right)
-\left(\frac{M}{n_\text{ex}}-N_c\right)\ln\left(1-\frac{n_\text{ex}N_c}{M}\right)
\nonumber\\[2ex]
&=&-\frac{V_m}{v_p}\left[\frac{\phi_c}{n}\ln\left(\frac{\phi_c}{n}\right)
+\frac{1-\phi_c'}{n_\text{ex}}\ln(1-\phi_c')\right],
\label{Sc0}
\end{eqnarray}
where $\phi_c'=(n_\text{ex}/n)\phi_c$ is the fraction of volume excluded by crowders to one another,
which exceeds the volume fraction $\phi_c$ excluded to polymer and solvent.
 
Going a step further, we approximately account for correlations between pairs of crowders due to overlap 
of crowder excluded-volume spheres by subtracting from the number of sites excluded to a crowder an estimate 
of the mean number of overlap sites (which otherwise would be doubly-counted). Still assuming a 
sufficiently dilute crowder solution that multi-particle clusters contribute negligibly 
to the microstate count, thus avoiding the complication of triplets and many-particle clusters of crowders,
the number of ways of placing $N_c$ indistinguishable crowders now can be written as
\begin{eqnarray}
&&
\hspace*{-0.4cm}
\Omega_c=\frac{M}{N_c!}(M-n_\text{ex})(M-2n_\text{ex}+\bar n_\text{ov})(M-3n_\text{ex}+3\bar n_\text{ov})
\nonumber\\[2ex]
&&
\hspace*{0.4cm}
\cdots\left[M-(N_c-1)n_\text{ex}+\frac{N_c(N_c-1)}{2}\bar n_\text{ov}\right],
\label{Omegac}
\end{eqnarray}
where $\bar n_\text{ov}$ is the mean number of sites in the pair overlap regions of crowder 
excluded-volume spheres, averaged over crowder pair configurations.
Again applying the Stirling approximation, we have
\begin{eqnarray}
\ln\Omega_c &\simeq& -N_c\ln\left(\frac{N_c}{M}\right) + N_c 
\nonumber\\[2ex]
&+& \int_0^{N_c}dm\,\ln\left(1-\frac{n_\text{ex}}{M}m+\frac{\bar n_\text{ov}}{2M}m^2\right).
\label{lnOmegac1}
\end{eqnarray}
We can calculate $\bar n_\text{ov}$ by integrating the intersection volume $v_\text{ov}(r)$ of two 
excluded-volume spheres, of radius $2R_c$ and center-center separation $r$, over the overlap volume, 
being a spherical shell of inner radius $2R_c$ and outer radius $4R_c$, and multiplying by the 
density of sites $1/v_p$ and the probability $1/M$ of a site being occupied:
\begin{equation}
\bar n_\text{ov}=\frac{4\pi}{Mv_p^2}\int_{2R_c}^{4R_c} dr\, r^2 v_\text{ov}(r),
\label{nov-integral}
\end{equation}
where $v_\text{ov}(r)=(\pi/12)(8R_c+r)(4R_c-r)^2$. The exact result,
$\bar n_\text{ov}=(17/32)n_\text{ex}^2/M$, is closely approximated by
$\bar n_\text{ov}\simeq n_\text{ex}^2/(2M)$.
Remarkably (or fortuitously), this approximation allows completing the square in the argument of the 
logarithm on the right side of Eq.~(\ref{lnOmegac1}), greatly simplifying the integral and leading to
\begin{equation}
\ln\Omega_c = -N_c\ln\left(\frac{N_c}{M}\right)+N_c+2\int_0^{N_c}dm\,
\ln\left(1-\frac{n_\text{ex}}{2M}m\right).
\label{lnOmegac2}
\end{equation}
Evaluating the remaining integral, we obtain for the crowder entropy
%
\begin{eqnarray}
\frac{S_c}{k_B}&=&-\frac{V_m}{v_p}\left[\frac{\phi_c}{n}\ln\left(\frac{\phi_c}{n}\right)+\frac{\phi_c}{n}\right.
\nonumber\\[2ex]
&+&\left.\frac{4}{n_\text{ex}}\left(1-\frac{\phi_c'}{2}\right)\ln\left(1-\frac{\phi_c'}{2}\right)\right],
\label{Sc1}
\end{eqnarray}
to be compared with Eq.~(\ref{Sc0}), which neglects pair correlations.
Modeling the crowders as hard spheres, each excluding eight times its own volume to others,
we take $n_\text{ex}=8n$ and $\phi_c'=8\phi_c$, finally yielding
\begin{equation}
\frac{S_c}{k_B}=-\frac{V_m}{v_p}\left[\frac{\phi_c}{n}\ln\left(\frac{\phi_c}{n}\right)+\frac{\phi_c}{n}
+\frac{1-4\phi_c}{2n}\ln(1-4\phi_c)\right].
\label{Sc2}
\end{equation}

The associated crowder free energy density is given by
\begin{equation}
\beta\frac{F_c}{V_m} = \rho_c\ln(\rho_cv_p)+\rho_c+\frac{1-4\phi_c}{2v_c}\ln\left(1-4\phi_c\right),
\label{fc}
\end{equation}
where $\beta=1/(k_BT)$ and $\rho_c=\phi_c/v_c$ is the crowder number density. 
Comparing the low-density expansion,
\begin{equation}
\beta\frac{F_c}{V_m} = \rho_c\ln(\rho_cv_p)-\rho_c+4v_c\rho_c^2+\frac{16}{3}v_c^2\rho_c^3+\cdots,
\label{fc-expansion}
\end{equation}
with the corresponding virial expansion~\cite{HansenMcDonald} for a fluid of hard spheres of volume $v_c$,
\begin{equation}
\beta\frac{F_c}{V_m} = \rho_c\ln(\rho_cv_p)-\rho_c+B_2v_c\rho_c^2+\frac{1}{2}B_3v_c^2\rho_c^3+\cdots,
\label{f-virial}
\end{equation}
we find that our approximate free energy gives the correct ideal-gas free energy density,
$\beta F_\text{id}/V_m=\rho_c\ln(\rho_cv_p)-\rho_c$, the exact hard-sphere second virial coefficient, 
$B_2=4$, and a third virial coefficient, $B_3=32/3$, in very close agreement with the hard-sphere 
value of $B_3=10$.

In passing, we note that our approach for incorporating excluded volume in a lattice model of a solution 
of crowder particles may be compared with the theory of Marques Fernandes \etalia~\cite{fernandes-levin-arenzon2007}
for modeling the thermodynamics of a lattice gas of hard cubes, each occupying $n$ sites. 
When translated into our symbols, the latter theory (with $v_c=\lambda^d$) predicts a free energy density
\begin{equation}
\beta\frac{F_c}{V_m} = \rho_c\ln\phi_c+\frac{1+\ln n}{v_c}(1-\phi_c)\ln(1-\phi_c).
\label{fc-fernandes}
\end{equation}
Since it is constructed to interpolate between the low- and high-density limits, Eq.~(\ref{fc-fernandes}) 
yields the correct ideal-gas free energy (as does our theory). In contrast with our approach, however, 
the theory of Marques Fernandes \etalia applies at all densities, while our theory focuses on and is limited 
to dilute crowder concentrations.

Considering now the polymer monomers and solvent molecules, for a given configuration of $N_c$ crowders
that exclude $nN_c$ sites, the number of distinct arrangements of these other components is given by
\begin{equation}
\Omega_{ps} = \frac{(M-n N_c)!}{N_s!N_p!}.
\label{Omegaps}
\end{equation}
The associated polymer-solvent entropy is given by
\begin{eqnarray}
\frac{S_{ps}}{k_B} &=& N_s\ln\left(\frac{M-n N_c}{N_s}\right)-N_p\ln N_p
\nonumber\\[2ex]
&=&-\frac{V_m}{v_p}\left[
\tilde\phi_s\ln\tilde\phi_s-(1-\phi_c)\ln(1-\phi_c)+\phi_p\ln\phi_p\right],
\nonumber\\[1ex]
\label{Sps}
\end{eqnarray}
where we have invoked the volume fraction sum rule [Eq.~(\ref{phisum})]. The last term on the right side 
represents the entropy of the polymer monomers if considered as independent particles. However, 
connecting the monomers into chains and cross-linking the chains into a network reduces the polymer entropy 
by a factor of one over the microgel molecular weight $N_\text{mon}$~\cite{doi2013soft,deGennes1979}. 
Assuming that $N_\text{mon}\gg 1$, the polymer contribution is then completely negligible.
Combining Eqs.~(\ref{Sc2}) and (\ref{Sps}), the final mixing entropy takes the form
\begin{eqnarray}
\frac{S_\text{mix}}{k_B} &=& -\frac{V_m}{v_p}\left[
\tilde\phi_s\ln\tilde\phi_s-(1-\phi_c)\ln(1-\phi_c)
\right.
\nonumber\\[2ex]
&+& \left.
\frac{\phi_c}{n}\ln\left(\frac{\phi_c}{n}\right)
+\frac{\phi_c}{n}+\frac{1-4\phi_c}{2n}\ln(1-4\phi_c)\right].
\nonumber\\[1ex]
\label{Smix}
\end{eqnarray}
The form of the last term on the right side makes clear that our model is limited to the range 
$\phi_c<0.25$, such that the argument of the logarithm remains positive. 
Note that in the limit in which the crowders shrink to a size comparable to that of monomers, 
and then occupy only single sites ($n_\text{ex}=n=1$), the crowders would no longer exclude 
extra volume. In this case, the crowder entropy would be described by Eq.~(\ref{Sc0}) and 
the mixing entropy would reduce simply to
\begin{equation}
\frac{S_\text{mix}}{k_B} = -N_s\ln\tilde\phi_s-N_c\ln\phi_c \quad (n_\text{ex}=n=1).
\label{Smix_limit}
\end{equation}

In the Flory-Huggins lattice model, which includes only nearest-neighbor interactions, 
the mean energy of the mixture can be expressed as a sum of pair energies,
\begin{eqnarray}
\bar{E}_\text{mix} &=& \epsilon_{pp}\bar{N}_{pp}+\epsilon_{ss}\bar{N}_{ss}+\epsilon_{cc}\bar{N}_{cc}
\nonumber\\[2ex]
&+&\epsilon_{ps}\bar{N}_{ps}+\epsilon_{pc}\bar{N}_{pc}+\epsilon_{sc}\bar{N}_{sc},
\label{Emean}
\end{eqnarray}
where $\epsilon_{ij}$ is the interaction energy between nearest-neighbor particles of species 
$i$ and $j$ and $\bar{N}_{ij}$ is the mean number of $ij$ nearest-neighbor pairs. In the associated
mean-field theory, in which correlations are neglected and the neighboring sites of each particle 
are simply replaced by average occupation fractions for the whole lattice, $\bar{N}_{ij}$ is 
approximated by $z_{ij}N_i N_j/M$, where $z_{ij}$ is the appropriate coordination number,
i.e., number of $j$ particles that can be neighbors of an $i$ particle. The mean mixing energy 
$E_\text{mix}$, defined as the difference in mean energy of the mixture and the individual 
pure components, is then approximated as
\begin{eqnarray}
\beta\bar{E}_\text{mix} &=& M\left(\chi_{ps}\phi_p\tilde\phi_s
+\chi_{pc}\phi_p\phi_c+\chi_{sc}\tilde\phi_s\phi_c\right)
\nonumber\\[2ex]
&=&\chi_{ps}N_p\tilde\phi_s+\chi_{pc}N_p\phi_c+\chi_{sc}N_s\phi_c,
\label{Emix}
\end{eqnarray}
where the Flory solvency parameters, $\chi_{ps}$, $\chi_{pc}$, and $\chi_{sc}$, are combinations 
of the $\epsilon_{ij}$ and $z_{ij}$ parameters. 
As an example, the polymer-solvent parameter is given by 
\begin{equation}
\chi_{ps} = z_{ps}\left(\epsilon_{ps}-\frac{\epsilon_{pp}+\epsilon_{ss}}{2}\right).
\label{chi_ps}
\end{equation}
Note that, because crowders occupy multiple sites, their associated coordination numbers 
($z_{pc}$, $z_{sc}$, $z_{cc}$) exceed those involving only polymer or solvent, 
although these factors are ultimately absorbed into the solvency parameters.
The mixing free energy is the sum of the mixing entropy and mixing energy.

A further contribution to the free energy comes from the conformational entropy of the 
polymer chains and the freedom of a polymeric gel to swell and deswell. As in our previous work
~\cite{urich2016swelling,weyer2018concentration}, we model this contribution using the mean-field 
Flory-Rehner theory of polymer networks~\cite{Flory-Rehner1943-I,Flory-Rehner1943-II,flory1953principles}, 
which treats the polymers as Gaussian coils, assumes affine swelling of the cross-linked network,
and predicts an elastic free energy,
\begin{equation}
\beta F_\text{el} = \frac{3}{2}N_\text{ch}(\alpha^2-1-\ln\alpha).
\label{Fel}
\end{equation}
While the excluded volume of crowders may certainly affect the swelling ratio in
equilibrium, we assume that the basic functional form of $F_\text{el}$ remains unchanged. 

The final contribution to the free energy of a polymeric gel in a solution of crowders 
comes from the entropy loss resulting from penetrating a polymer network by a crowder.
Modeling the polymer chains as fluctuating ellipsoids whose principal radii follow the statistics
of a random walk~\cite{kuhn1934shape,rudnick1987shapes,murat1998many,eurich2001soft,
lim-denton-JCP2014,lim-denton-JCP2016,lim-denton-SM2016,davis-dentonJCP2018},
consistent with the Gaussian approximation of the Flory-Rehner theory, 
and assuming the crowders are small enough as to perturb the conformation of only one 
chain at a time, we estimate the penetration free energy from 
polymer field theory~\cite{eisenriegler1996,hanke1999,eisenriegler2000,eisenriegler2003} 
as the entropic cost of inserting a hard sphere into a polymer coil in a good solvent:
\begin{equation}
\beta\varepsilon_\text{pen}(\alpha) = \frac{4.416}{q^{1.29932}},
\label{epsilon-pen}
\end{equation}
where $q=(\alpha/\alpha_{\rm res})R_g/R_c$ is the polymer-to-crowder size ratio in a gel of 
swelling ratio $\alpha$ relative to the reservoir swelling ratio $\alpha_{\rm res}$ in pure solvent.
Note that this definition of the size ratio, which incorporates the effect of swelling/deswelling
on the mean polymer size, is consistent with the affine swelling assumption.
In the limit $n\to 1$, in which the crowders become indistinguishable from solvent molecules,
$q\gg 1$ and we can set $\varepsilon_\text{pen}=0$.

In the current modified Flory-Rehner approximation, the total Helmholtz free energy of a gel 
in the solution is finally given by
\begin{equation}
F = \bar{E}_\text{mix}-TS_\text{mix}+F_\text{el}+N_c\varepsilon_\text{pen},
\label{Ftot}
\end{equation}
where $N_c$ here represents the number of crowders inside the gel. We emphasize that, although
microscopic species are represented by density fields in this coarse-grained level of modeling, 
microgels in a suspension must be rendered explicitly, as discussed below.

In equilibrium, the chemical potentials of solvent and crowders inside the gel, $\mu_s$ and $\mu_c$,
must equal the respective chemical potentials outside, $\mu^0_s$ and $\mu^0_c$, which are fixed 
via exchange of both species with the reservoir: 
\begin{equation}
\mu_s(\alpha, \phi_c, \phi_c^0) = \mu_s^0(\phi_c^0),
\quad
\mu_c(\alpha, \phi_c, \phi_c^0) = \mu_c^0(\phi_c^0).
\label{equilibrium}
\end{equation}
These relations constitute a system of two nonlinear equations, which, for a given reservoir
crowder concentration $\phi_c^0$, can be numerically solved for the two unknowns, namely, 
the crowder and polymer concentrations inside the gel, $\phi_c$ and $\phi_p$
(or equivalently the swelling ratio $\alpha$). Given $\phi_c$ and $\phi_p$, the solvent 
concentration inside the gel $\phi_s$ follows from Eq.~(\ref{phisum}).

The solvent and crowder chemical potentials in Eqs.~(\ref{equilibrium}) are derived from 
the total free energy:
\begin{equation}
\mu_s = \left(\frac{\partial F}{\partial N_s}\right)_{N_p, N_c},
\quad
\mu_c = \left(\frac{\partial F}{\partial N_c}\right)_{N_p, N_s},
\label{mus-muc}
\end{equation}
where the partial derivatives are evaluated subject to Eqs.~(\ref{M})-(\ref{phisum}), but allowing 
for a change of volume of the gel (and thus number of lattice sites) as solvent and crowders are 
absorbed or expelled~\cite{doi2013soft}. The separation of the free energy into mixing, elastic, 
and penetration terms implies a corresponding separation of the chemical potentials.
Substituting the mixing free energy [Eqs.~(\ref{Smix}) and (\ref{Emix})] into Eq.~(\ref{mus-muc}), 
we obtain the mixing contributions to the solvent and crowder chemical potentials:
\begin{eqnarray}
\beta\mu_{s,\text{mix}} &=& \ln\tilde\phi_s+\phi_p+\frac{\phi_c}{n}-\ln(1-\phi_c)+\frac{1}{2n}\ln(1-4\phi_c)
\nonumber\\[2ex]
&+&\chi_{ps}\phi_p(1-\tilde\phi_s)-\chi_{pc}\phi_p\phi_c+\chi_{sc}\phi_c(1-\tilde\phi_s)
\nonumber\\[1ex]
\label{mus-mix}
\end{eqnarray}
and
\begin{eqnarray}
\beta\mu_{c,\text{mix}} &=& \ln\left(\frac{\phi_c}{n}\right)+\phi_c+n\phi_p-\frac{3}{2}\ln(1-4\phi_c)
\nonumber\\[2ex] 
&-&n[\chi_{ps}\phi_p\tilde\phi_s-\chi_{pc}\phi_p(1-\phi_c)-\chi_{sc}\tilde\phi_s(1-\phi_c)].
\nonumber\\[1ex]
\label{muc-mix}
\end{eqnarray}
From the elastic free energy [Eq.~(\ref{Fel})], we obtain the elastic contributions:
\begin{equation}
\beta\mu_{s,\text{el}} = \frac{N_\text{ch}}{N_s}(\alpha^2-1/2)\phi_s
\label{mus-el}
\end{equation}
and
\begin{equation}
\beta\mu_{c,\text{el}} = \frac{N_\text{ch}}{N_c}(\alpha^2-1/2)\phi_c,
\label{muc-el}
\end{equation}
which result from [Eq.~(\ref{Vm})] and the dependence of the swelling ratio $\alpha$ on volume $V_m$.
Finally, from the penetration free energy [Eq.~(\ref{epsilon-pen})], we obtain the 
penetration contributions:
\begin{equation}
\mu_{s,\text{pen}} = -\frac{\gamma}{3}\frac{v_s}{v_c}\phi_c\varepsilon_\text{pen},
\quad
\mu_{c,\text{pen}} = \left(1-\frac{\gamma}{3}\phi_c\right)\varepsilon_\text{pen}.
\label{mu-pen}
\end{equation}

From Eqs.~(\ref{mus-mix}) and (\ref{muc-mix}), the corresponding solvent and crowder
chemical potentials in the reservoir (absent polymer) are given by
\begin{equation}
\mu_s^0 = \frac{\phi_c^0}{n}+\frac{1}{2n}\ln(1-4\phi_c^0)+\chi_{sc}(\phi_c^0)^2
\label{mus0}
\end{equation}
and
\begin{equation} 
\mu_c^0 = \ln\left(\frac{\phi_c^0}{n}\right)+\phi_c^0-\frac{3}{2}\ln(1-4\phi_c^0)+\chi_{sc}n(1-\phi_c^0)^2.
\label{muc0}
\end{equation}
Substituting Eqs.~(\ref{mus-mix})-(\ref{muc0}) into Eqs.~(\ref{equilibrium}), we can solve for
the equilibrium values of $\alpha$ and $\phi_c$ for a polymeric gel immersed in a solution 
of crowders of given concentration $\phi_c^0$. Alternatively, and equivalently, 
we can minimize with respect to variations in $N_s$ and $N_c$ the appropriate thermodynamic potential 
in the semigrand canonical ensemble, namely, the Landau free energy, 
\begin{equation}
L=F-\mu^0_s N_s-\mu^0_c N_c,
\label{L}
\end{equation}
a Legendre transformation of the Helmholtz free energy.

For a suspension of $N_m$ interacting microgels, the total free energy includes contributions 
accounting for the entropy associated with microgel center-of-mass configurations and the 
total energy of interparticle interactions. Since microgels have polymeric degrees of freedom,
the interaction energy depends on the internal states of the particles and may have many-body 
contributions. Assuming strictly pairwise interactions, however, and preaveraging over most internal 
degrees of freedom, we approximate the interaction energy by a pairwise sum
\begin{equation}
U_\text{H}=\sum_{i=1}^{N_m}\sum_{j>i=1}^{N_m}v_\text{H}(r_{ij}),
\label{UH}
\end{equation}
where
\begin{equation}
v_\text{H}(r_{ij})=B_{ij}\left(1-\frac{r_{ij}}{a_i+a_j}\right)^{5/2}
\label{vH}
\end{equation}
is the Hertz elastic potential between microgels $i$ and $j$ with swollen radii $a_i$ and $a_j$ 
and center-to-center separation $r_{ij}$.
The amplitude depends on the particle radii, Young's moduli $Y_i$, and Poisson's ratios $\nu_i$ 
via~\cite{landau-lifshitz1986}
\begin{equation}
B_{ij}=\frac{8}{15}\left(\frac{1-\nu_i^2}{Y_i}+\frac{1-\nu_j^2}{Y_j}\right)^{-1}
(a_i+a_j)^2\sqrt{a_i a_j}.
\label{Bij}
\end{equation}
Since the Young's modulus of a gel is proportional to $k_BT$ and the 
cross-link density~\cite{deGennes1979}, we take
\begin{equation}
\beta B_{ij}=Y_c N_{\rm ch}\frac{(a_i+a_j)^2\sqrt{a_i a_j}}{a_i^3+a_j^3},
\label{Bij2}
\end{equation}
where $Y_c$ is a dimensionless factor introduced to calibrate the model with experiment.
We note that heterogeneously cross-linked microgels can be more accurately described by 
a multi-Hertzian model~\cite{zaccarelli-multiHertzian2018,moncho-jorda2025} that accounts for 
different cross-linker densities in the core and outer shells.

To determine the equilibrium state of a suspension of microgels interacting via Hertz pair potentials,
we perform Monte Carlo simulations with trial moves combining displacements, size changes, and 
crowder concentration changes of all microgels. The acceptance probabilities for the trial moves 
can be derived from the Hamiltonian $H$ of the system and the partition function,
\begin{equation}
{\cal Z}=\la\la\la e^{-\beta H}\ra_m\ra_p\ra_c,
\label{Z1}
\end{equation}
where $\la~\ra_m$, $\la~\ra_p$, and $\la~\ra_c$ denote respective traces over microgel center-of-mass,
polymeric, and crowder degrees of freedom, the latter governing partitioning of crowders between 
inside and outside the microgels. Neglecting many-body interactions, the Hamiltonian separates 
into one-body and pair terms, $H_0$ and $H_{\rm pair}$:
\begin{equation}
H=H_0+H_{\rm pair}.
\label{H}
\end{equation}

Assuming that, of all the polymeric degrees of freedom, swelling is governed by a relatively small
subset (denoted collectively by $a$), and denoting the remaining degrees of freedom by $p'$, 
we may write the partition function as
\begin{equation}
{\cal Z}=\la\la\la e^{-\beta H_{\rm eff}}\ra_m\ra_a\ra_c,
\label{Z2}
\end{equation}
where $\la~\ra_a$ and $\la~\ra_c$ denote averages over all microgel radii and 
interior numbers of crowders and
\begin{equation}
H_{\rm eff}=-k_BT\ln\la e^{-\beta(H_0+H_{\rm pair})}\ra_{p'}
\label{Heff}
\end{equation}
is an {\it effective} Hamiltonian resulting from the trace $\la~\ra_{p'}$ over $p'$ coordinates.
Now making a mean-field approximation that neglects correlations between microgel swelling and
pair interactions, and thus decouples one-body (single-microgel) and pair coordinates, we write
\begin{eqnarray}
H_{\rm eff}&=&-k_BT\ln\left(\la e^{-\beta H_0}\ra_{p'}\la e^{-\beta H_{\rm pair}}\ra_{p'}\right)
\nonumber\\[2ex]
&=&F_0-k_BT\ln\la e^{-\beta U_{\rm eff}}\ra_{p'}.
\label{Heff2}
\end{eqnarray}
Here we define a single-microgel polymeric free energy,
\begin{equation}
F_0(a, \phi_c)=-k_BT\ln\la e^{-\beta H_0}\ra_{p'}
=\sum_{i=1}^{N_m}F(a_i, \phi_{c,i}),
\label{Fm}
\end{equation}
which depends only on the swollen microgel radii and interior crowder volume fractions,
and an effective pair interaction energy,
\begin{equation}
U_{\rm eff}(a)=-k_BT\ln\la e^{-\beta H_{\rm pair}}\ra_{p'},
\label{Ueff}
\end{equation}
which depends on the swollen radii and center-of-mass coordinates of the microgels. 
We associate $F_0$ with the Flory-Rehner free energy in Eq.~(\ref{Ftot}), summed over
all microgels, and $U_{\rm eff}$ with the Hertz potential energy in Eq.~(\ref{UH}). 
After tracing over a subset of internal degrees of freedom of the microgels, the resulting 
canonical partition function in this coarse-grained model reduces to
\begin{equation}
{\cal Z}=\la\la e^{-\beta F_0(a)}\la e^{-\beta U_\text{H}(a)}\ra_m\ra_a\ra_c.
\label{Z3}
\end{equation}
In the semigrand canonical ensemble, which allows for exchange of solvent and crowders
with a reservoir solution of crowders, the semigrand partition function then takes the form
\begin{equation}
\Xi=\la\la e^{-\beta L_0(a, \phi_c)}\la e^{-\beta U_\text{H}(a)}\ra_m\ra_a\ra_c,
\label{Xi}
\end{equation}
where $L_0$ represents a sum over all microgels of the single-microgel Landau free energy of Eq.~(\ref{L}),
\begin{equation}
L_0(a, \phi_c) = \sum_{i=1}^{N_m}L(a_i, \phi_{c, i}),
\label{L0}
\end{equation}
which depends on the radii and interior crowder volume fractions of all microgels.

\section{Computational Methods}\label{methods}
To predict the equilibrium state, i.e., mean swelling ratio and interior crowder volume fraction, 
of a polymeric gel immersed in a solution of crowders, we numerically solved the coupled 
nonlinear Equations~(\ref{equilibrium}), equating chemical potentials inside and outside the gel. 
In practice, we used the fsolve function from the optimize module in the SciPy library of Python 
to simultaneously solve for $\alpha$ and $\phi_c$, looping over $\phi_c^0$ and using the 
solution for each iteration as an initial estimate for the next.

As a consistency check on our numerical computations, as well as to extend our study to microgel 
suspensions, we also performed Monte Carlo (MC) simulations~\cite{frenkel-smit2001,binder2010,
landau2021guide} of either a single microgel or many interacting microgels. We emphasize that, 
while our model includes explicit microgel particles, the polymer, solvent, and crowder components 
are implicit, entering only as densities in the one-body Landau free energy [Eq.~(\ref{L0})]. 
The implicit reservoir solution of crowders fixes the chemical potentials, $\mu_s^0$ and $\mu_c^0$, 
that appear in $L_0$.

\begin{figure}
\includegraphics[width=0.9\columnwidth,angle=0]{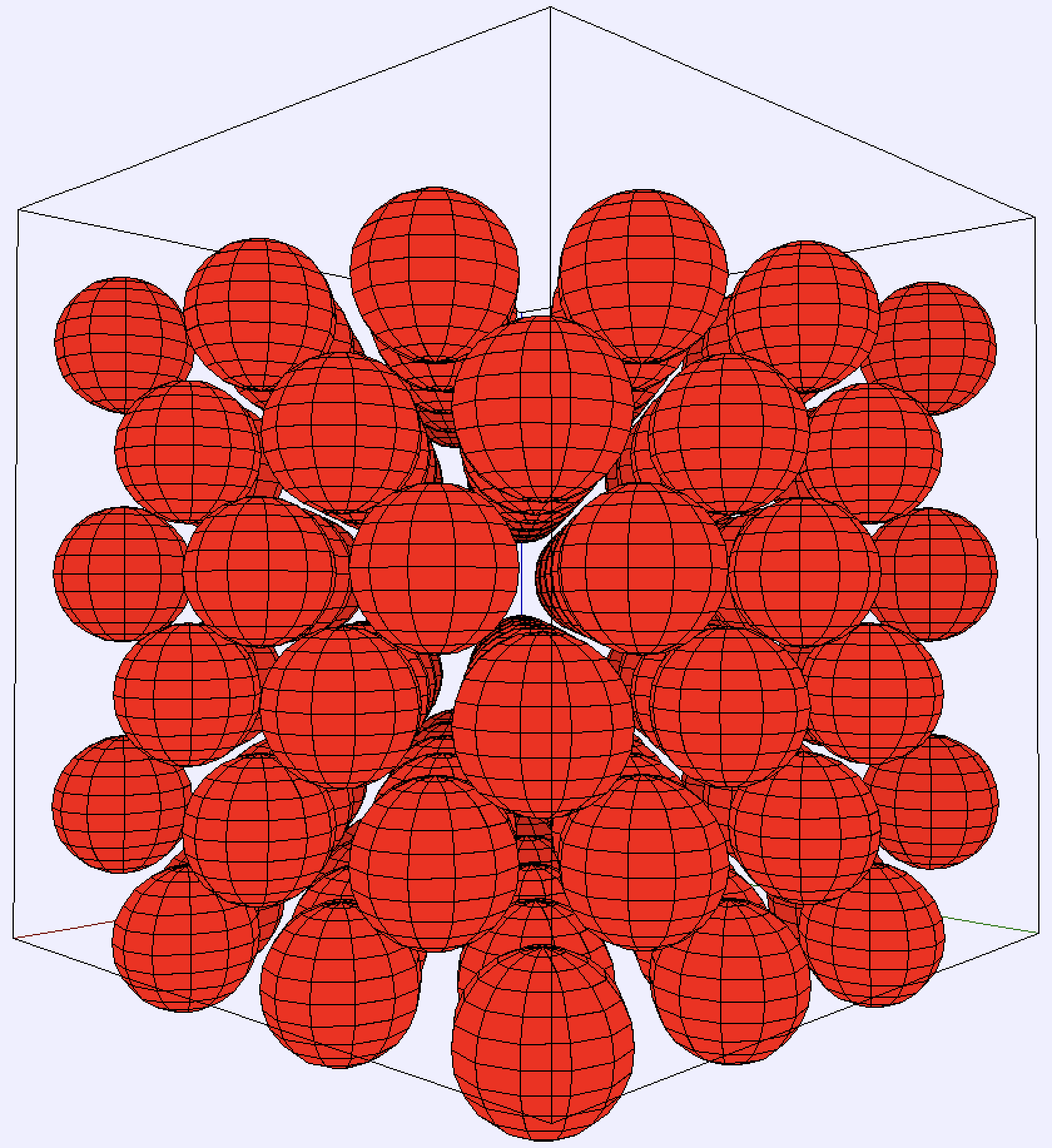}
\vspace*{-0.2cm}
\caption{Snapshot from an MC simulation of a suspension of microgels (red spheres) in an implicit 
solution of crowders. The microgels are confined to a cubic cell with periodic boundary conditions 
and initialized on an FCC lattice.}
\label{snapshot}
\end{figure}

A simulation proceeds through a series of trial moves, each comprising three independent random changes 
applied to every particle: a displacement of the microgel center (${\bf r}\to {\bf r}+\Delta {\bf r}$) 
and changes (increase or decrease) of the microgel radius ($a\to a+\Delta a$) and interior
crowder volume fraction ($\phi_c\to\phi_c + \Delta\phi_c$). While trial displacements are 
irrelevant for a single microgel, they affect the pair interaction energy for multiple microgels. 
The trial changes, $\Delta {\bf r}$, $\Delta a$, and $\Delta\phi_c$, are randomly generated within
tolerances set to optimize efficiency and accuracy.

A single MC step consists of attempting, then either accepting or rejecting, a trial move, 
in which the positions and internal states of all particles are changed. In the Metropolis 
algorithm~\cite{frenkel-smit2001,binder2010,landau2021guide}, the acceptance probability is determined 
by the ratio of the integrands of the partition function [Eq.~(\ref{Xi})] after and before the move:
\begin{equation}
P_{\rm acc}=\min\left\{e^{-\beta(\Delta L_0+\Delta U_\text{H})}, 1\right\},
\label{Pacc}
\end{equation}
where $\Delta L_0$ and $\Delta U_\text{H}$ are the resulting changes in the one-body Landau free energy 
[Eqs.~(\ref{L}) and (\ref{L0})] and the effective pair interaction energy [Eq.~(\ref{UH})], respectively.
Any trial move that decreases $L_0+U_\text{H}$ is automatically accepted, while a move that increases 
$L_0+U_\text{H}$ is accepted only with a probability equal to the Boltzmann factor for the change, 
$\Delta L_0+\Delta U_\text{H}$. In case of a rejection, the configuration is unchanged, but counts again 
in the statistics of the sampled ensemble. The Metropolis algorithm satisfies the condition of 
detailed balance, ensuring sampling from the correct semigrand canonical ensemble of configurations. 

It is through the acceptance probability [Eq.~(\ref{Pacc})] that the total Helmholtz free energy 
[Eq.~(\ref{Ftot})] predicted by our theory governs the evolution of the system toward equilibrium. 
Agreement between our simulations and numerical calculations (Sec.~\ref{results}) ultimately 
demonstrates that the Landau free energy is the appropriate thermodynamic potential. Furthermore, 
any inconsistency between the free energy and the chemical potentials of solvent and crowders 
would manifest in a discrepancy between results from the two methods, providing powerful
consistency and validation checks.

Figure~\ref{snapshot} shows a typical snapshot from a multi-particle simulation, in which the 
microgels are confined to a cubic cell with periodic boundary conditions applied to opposing faces. 
The particles are initialized on the sites of an FCC lattice, with each assigned an initial radius 
equal to its fully swollen radius in pure solvent and initial interior crowder volume fraction 
equal to the reservoir value.  
We custom-coded all simulations in Java within the Open Source Physics Library~\cite{osp-sip2007,osp2006}.

\section{Results and Discussion}\label{results}
Implementing the methods discussed in Sec.~\ref{methods}, we numerically solved Eqs.~(\ref{equilibrium})
and independently performed MC simulations to compute equilibrium swelling ratios and interior
crowder volume fractions of single microgels immersed in reservoir solutions of crowders of varying
concentration. For illustrative purposes, we chose system parameters as follows: 
microgel dry radius, $a_0=50$ nm, polymer monomer radius $R_{\rm mon}=0.3$ nm, segment length $l=0.6$ nm, 
and cross-link fraction $x=0.0005$. For suspensions, we chose the dry microgel volume fraction 
as $\phi_0=N_mV_0/V=0.03$. Although these parameters are representative of many experimental systems, 
our qualitative conclusions are insensitive to these choices. In their fully swollen state, such 
microgels have a mesh (pore) size, $l_m\simeq(v_p/x)^{1/3}\alpha$, and polymer chain radius of gyration 
[Eq.~(\ref{Rg})] both of roughly 15 nm.

For these particular system parameters, to ensure that the crowders are always smaller than the pores 
of a deswollen microgel, thus allowing for significant penetration of the network, we chose 
relatively small crowder radii $R_c=1.0, 1.25, 1.5$ nm, corresponding to crowder-to-monomer 
volume ratios $n=37, 72, 125$. In the case of larger pores, however, which are achievable by reducing the 
cross-link fraction or synthesizing the gel from polymers composed of longer segments, the crowders
may be correspondingly larger. It should be noted that our results for a single microgel can apply 
equally well to a bulk hydrogel having the same internal architecture (i.e., polymer type and cross-link 
fraction). Only for a suspension, in which interparticle interactions affect swelling and uptake, 
do the size and shape of the particles (spherical microgels) become relevant.

In our simulations, we first equilibrated the system for $10^5$ steps before collecting statistics 
over an additional $5\times 10^5$ steps, sampling at intervals of 100 steps, and computing averages 
to obtain mean values, $\la\alpha\ra$ and $\la\phi_c\ra$. As diagnostics, we computed on the fly 
the swelling ratio and chemical potentials to ensure that on approaching equilibrium
$\alpha\to{\rm constant}$, $\mu_s\to\mu_s^0$, and $\mu_c\to\mu_c^0$.

\begin{figure}
\includegraphics[width=\columnwidth,angle=0]{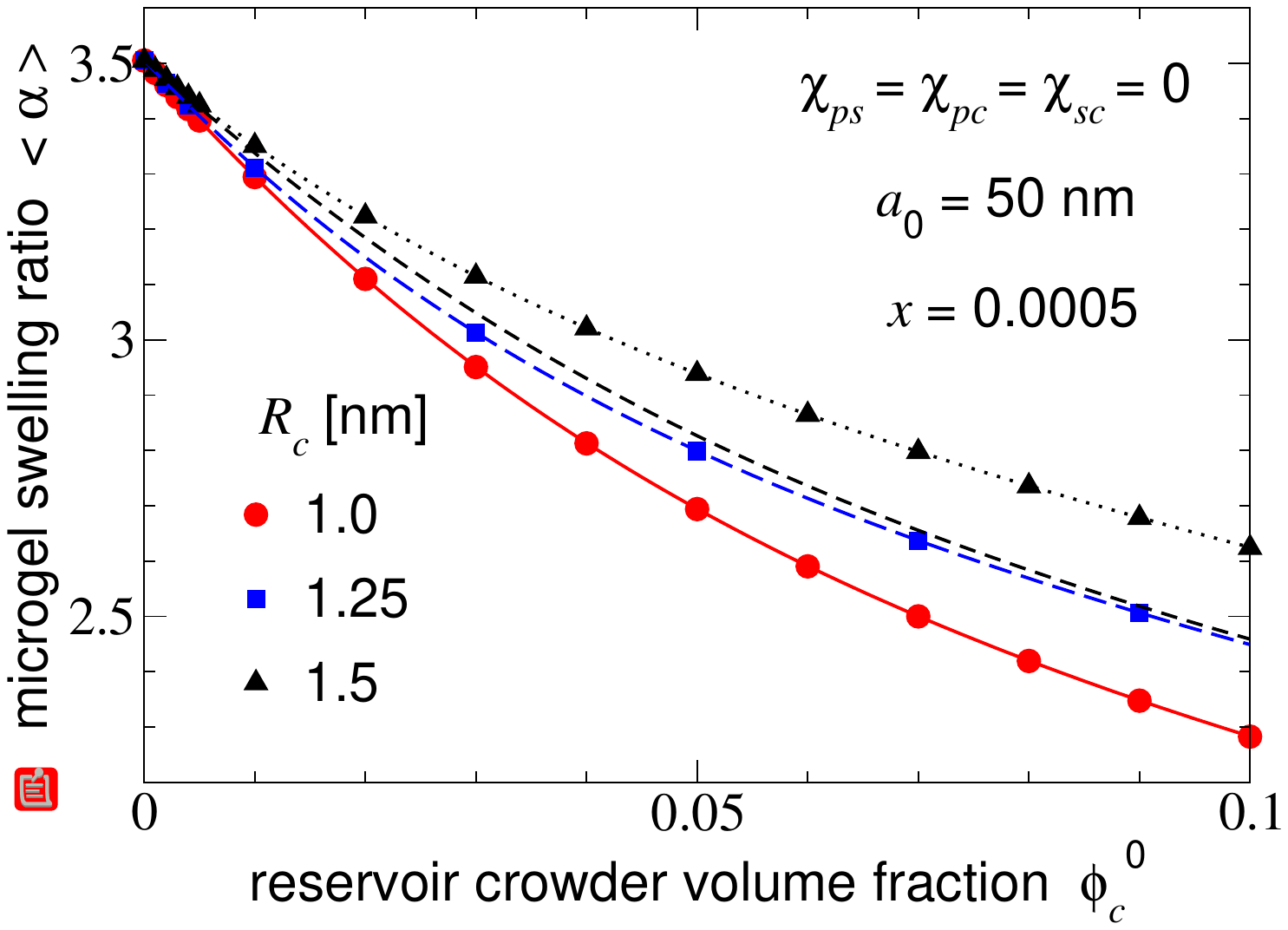}
\vspace*{-0.7cm}
\caption{Equilibrium mean swelling ratio $\la\alpha\ra$ of a bulk hydrogel or single microgel 
immersed in a solution of crowders versus reservoir volume fraction $\phi_c^0$ for crowder radii, 
$R_c=1.0, 1.25, 1.5$ nm. Other system parameters are microgel dry radius $a_0=50$ nm and 
cross-link fraction $x=0.0005$. 
Flory solvency parameters are all set to zero for good solvent conditions.
Numerical results (curves) and MC simulation data (symbols) are in near exact agreement. 
Black dashed curve is for $\varepsilon_\text{pen}=0$, $R_c=1.25$ nm.
}
\label{deswelling_Rc}
\end{figure}

\begin{figure}
\includegraphics[width=\columnwidth]{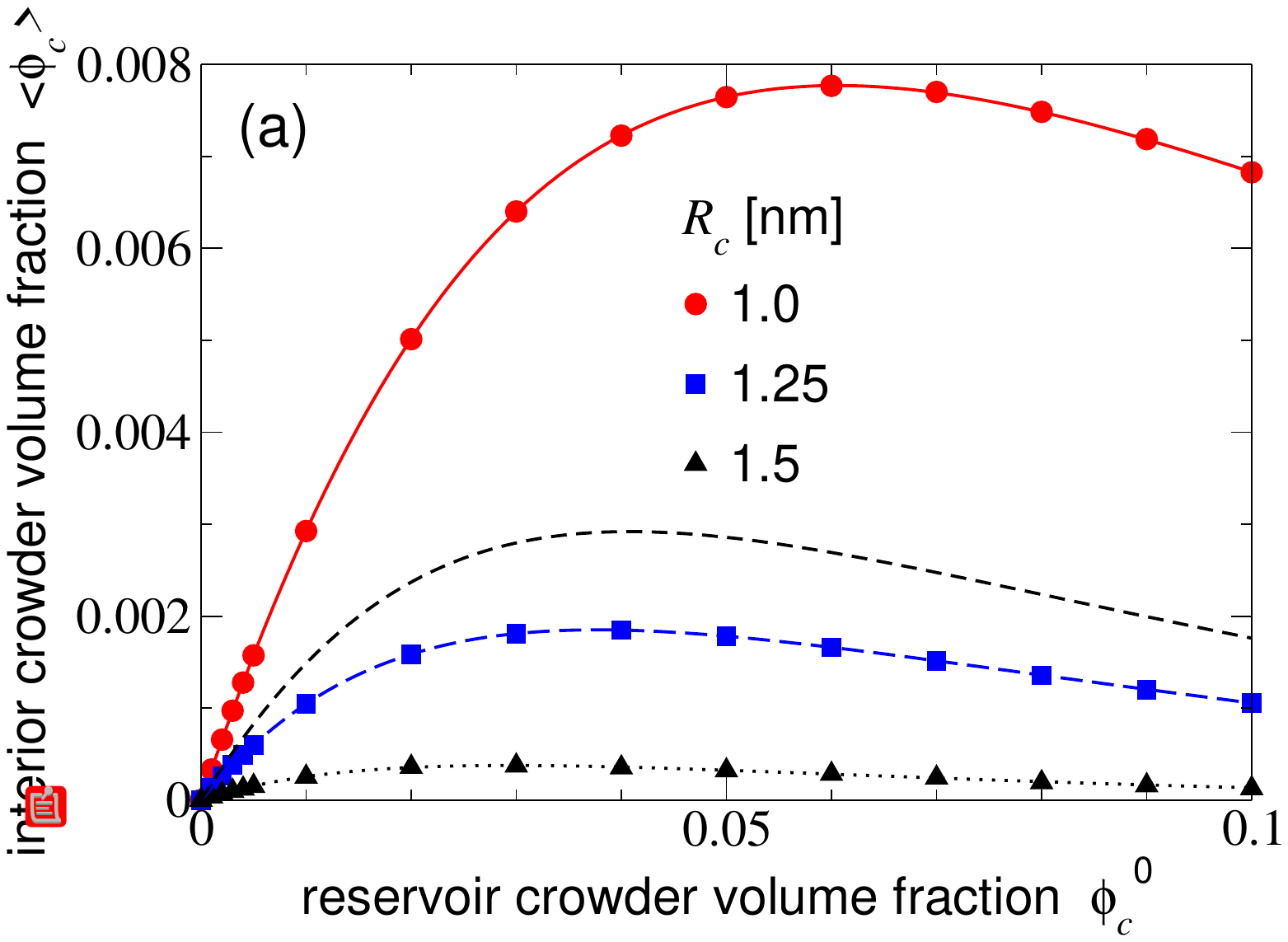}
\\[-7ex]
\includegraphics[width=\columnwidth]{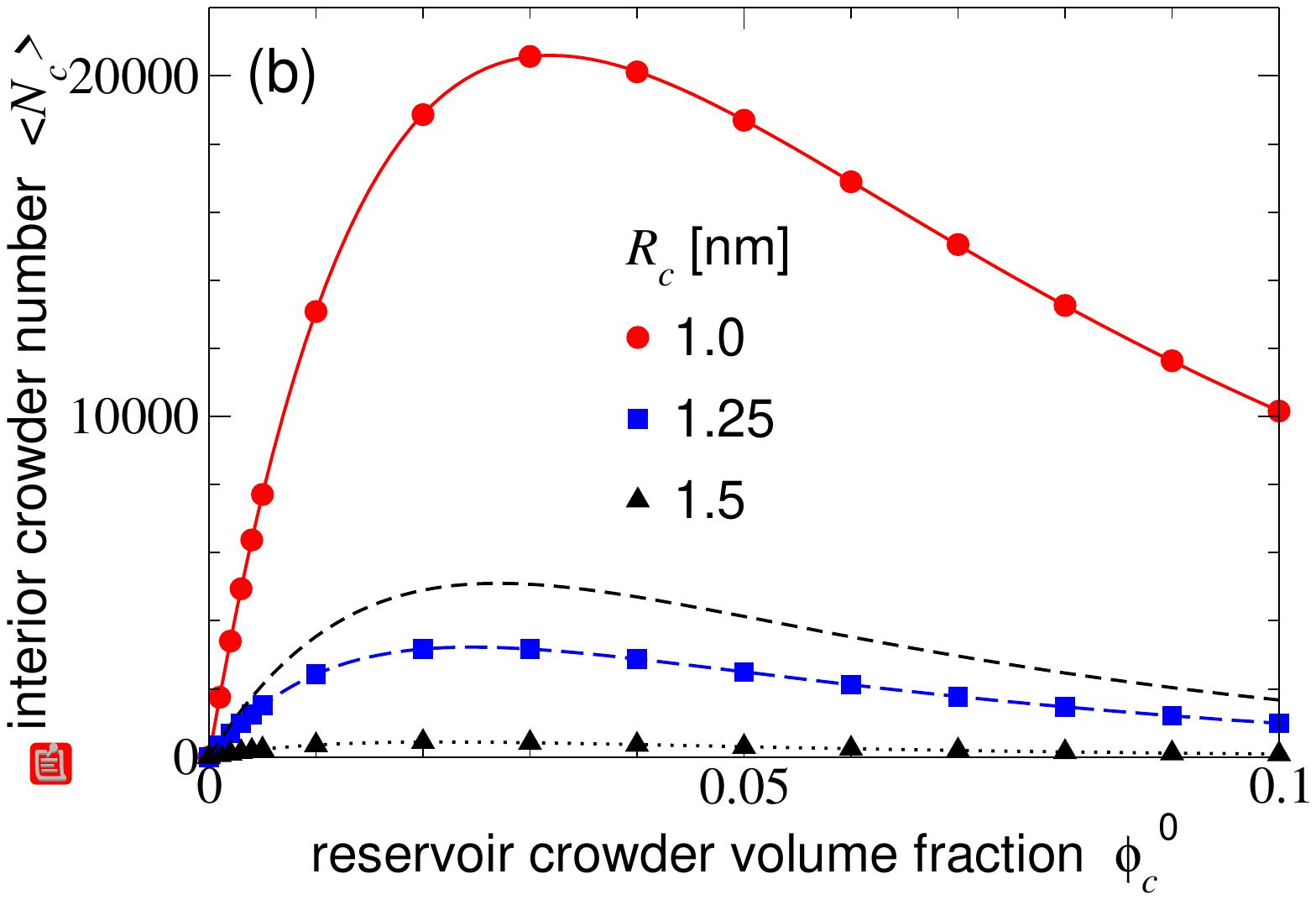}
\vspace*{-0.7cm}
\caption{Crowder uptake by a single microgel: interior crowder
(a) mean volume fraction $\la\phi_c\ra$ and (b) mean number $\la N_c\ra$ versus reservoir 
volume fraction $\phi_c^0$ for crowder radii, $R_c=1.0, 1.25, 1.5$ nm. 
Other system parameters are same as in Fig.~\ref{deswelling_Rc}. 
Numerical results (curves) and MC simulation data (symbols) are in near exact agreement. 
Black dashed curves are for $\varepsilon_\text{pen}=0$, $R_c=1.25$ nm.
}
\label{uptake_Rc}
\end{figure}

Figures~\ref{deswelling_Rc} and~\ref{uptake_Rc} show our predictions for the equilibrium swelling ratio 
and crowder uptake (volume fraction and number) of a bulk hydrogel, or single microgel, immersed in 
a solution of crowders as a function of crowder concentration in the reservoir. Initially assuming 
good solvent conditions, we set all of the Flory solvency ($\chi$) parameters to zero. Results are 
shown for three different crowder radii, roughly an order of magnitude larger than typical solvent 
molecules, yet small enough to be accommodated within the pores of the gel. To justify our neglect 
of crowder-crowder correlations, we restrict the reservoir volume fraction to $\phi_c^0\le 0.1$.
The near-exact agreement between our numerical calculations and independent simulation data validates 
our results and provides an important consistency check on our implementation of both methods.
Moreover, the experiments of Bradley \etalia~\cite{bradley-vincent2005} on PNIPAM microgels in 
solutions of poly(ethylene oxide) also demonstrate significant dependence of swelling and uptake 
on crowder (PEO) concentration and molecular weight. However, pH-dependent chemical (H-bonding) 
interactions between PEO and acrylic acid, as well as the freedom of PEO chains to change conformation,
complicate a direct comparison with predictions from our modeling.

We note that, within our current model, our numerical solution of Eqs.~(\ref{equilibrium}) and our
simulations always yield unique values of $\phi_s$ and $\phi_c$ (and thus a single $\alpha$ value), 
as confirmed by the observation that contour plots, with $\phi_c^0$ as parameter, exhibit only one 
crossing point. It is conceivable, however, that a model of microgels having a heterogeneous distribution 
of cross-linkers might exhibit two solutions, which would reflect an interesting bistable response to 
crowding.  Even if such microgels would have only one stable state, however, the interior distribution 
of crowders would be nonuniform. For example, core-shell microgels would likely harbor a lower concentration 
of crowders in the core, which is characterized by higher cross-link fraction and smaller pore size,
consistent with the trend seen in Fig.~\ref{uptake_x} below.

As seen in Fig.~\ref{deswelling_Rc}, with increasing reservoir crowder concentration, polymeric gels 
progressively deswell, the more so the smaller the crowders. This response to crowding is 
qualitatively akin to that observed in previous studies of linear homopolymers in a 
good solvent~\cite{davis-dentonJCP2018}, in which case a decrease in radius of gyration was attributed to 
the entropic cost of crowder penetration, as predicted by polymer field theory. A key difference, 
however, is that the polymers here are not isolated, but cross-linked into a network. 

Evidently, the entropic benefit to the network of absorbing solution and swelling, as described by the 
mixing part of the modified Flory-Rehner free energy [Eq.~(\ref{Ftot})], becomes suppressed by the 
entropic cost of accommodating the excluded volume of the crowders. The suppression increases with 
crowder size and concentration, consistent with the fact that the excluded-volume parameter $n$, 
which appears in the mixing entropy [Eq.~(\ref{Smix})] and chemical potentials [Eqs.~(\ref{mus-mix}) 
and (\ref{muc-mix})], varies with crowder size as $R_c^3$. While the penetration free energy contributes to 
deswelling, mixing entropy proves to be the dominant factor. In fact, setting $\varepsilon_\text{pen}=0$ 
only slightly reduces the deswelling (black dashed curve in Fig.~\ref{deswelling_Rc}).
From a mechanical perspective, by excluding volume, crowders modify the pressure of both the 
interior and exterior solutions, perturbing the delicate balance of osmotic pressure that determines 
the equilibrium swelling of a gel. From a solvency perspective, the crowders act as if an especially 
poor cosolvent for the polymer. The net result is that crowding-induced contraction of polymers 
in the network drives deswelling. 

As polymeric gels steadily deswell upon concentrating crowders in the reservoir solution, 
they also absorb crowders. As seen in Fig.~\ref{uptake_Rc}, with increasing $\phi_c^0$, as the 
chemical potential gradient drives crowders into the gel, the interior crowder volume fraction 
and number both initially increase. It now becomes apparent that the calculated numbers of crowders 
inside the gel are so high as to render explicit modeling computationally impractical, 
thus justifying our implicit model. 

Upon further increase in $\phi_c^0$, the number of crowders inside peaks and then {\it decreases},
as crowders are expelled into the surrounding solution. Similarly, the volume fraction inside 
reaches a maximum ($\phi_{c, \rm max}<\phi_c^0$), albeit at higher $\phi_c^0$ owing to the 
decrease in gel volume. Especially notable is the strong sensitivity to crowder size displayed 
in Fig.~\ref{uptake_Rc}. With increasing $R_c$, the uptake increases more gradually and peaks at 
lower values of $\phi_c^0$. This trend is again consistent with the excluded-volume factor $n$ 
in the mixing entropy varying as $R_c^3$. Thus, absorption of larger crowders is suppressed by 
excluded-volume penalties in both the penetration free energy and mixing entropy. 

The remarkable nonmonotonic response of polymeric gels to crowding reflects a compromise between mixing, 
elastic, and penetration contributions to the free energy. At lower reservoir concentrations, 
mixing entropy strongly favors uptake, while at higher concentrations, mixing entropy and 
penetration free energy both drive expulsion. Further analysis reveals that competition between 
the solvent and crowder terms in the mixing free energy dominates the uptake and swelling 
responses. Neglecting the penetration free energy (setting $\varepsilon_\text{pen}=0$) still leads to 
an uptake maximum, but at a higher concentration (black dashed curves in Fig.~\ref{uptake_Rc}).
We have also checked that omitting the swelling ratio factor $\alpha$ from the polymer-to-crowder 
size ratio $q$ in Eq.~(\ref{epsilon-pen}) would produce only a slight enhancement of swelling 
and uptake, without qualitatively changing the behavior. 

The observed swelling and uptake behavior highlights the complex nonlinear relationship between 
reservoir concentration and capacity for uptake by polymeric gels. Tuning this relationship may be 
important for engineering and controlling the uptake and release of nanoscale particles by such gels. 
We emphasize that our model of macromolecular crowding highlights the role of excluded volume 
in the mixing entropy [Eq.~(\ref{Smix})]. However, attractive or repulsive interactions between 
crowders and the gel can be incorporated through the Flory solvency parameters ($\chi_{ij}$) 
in the mean mixing energy [Eq.~(\ref{Emix})].

\begin{figure}
\includegraphics[width=\columnwidth,angle=0]{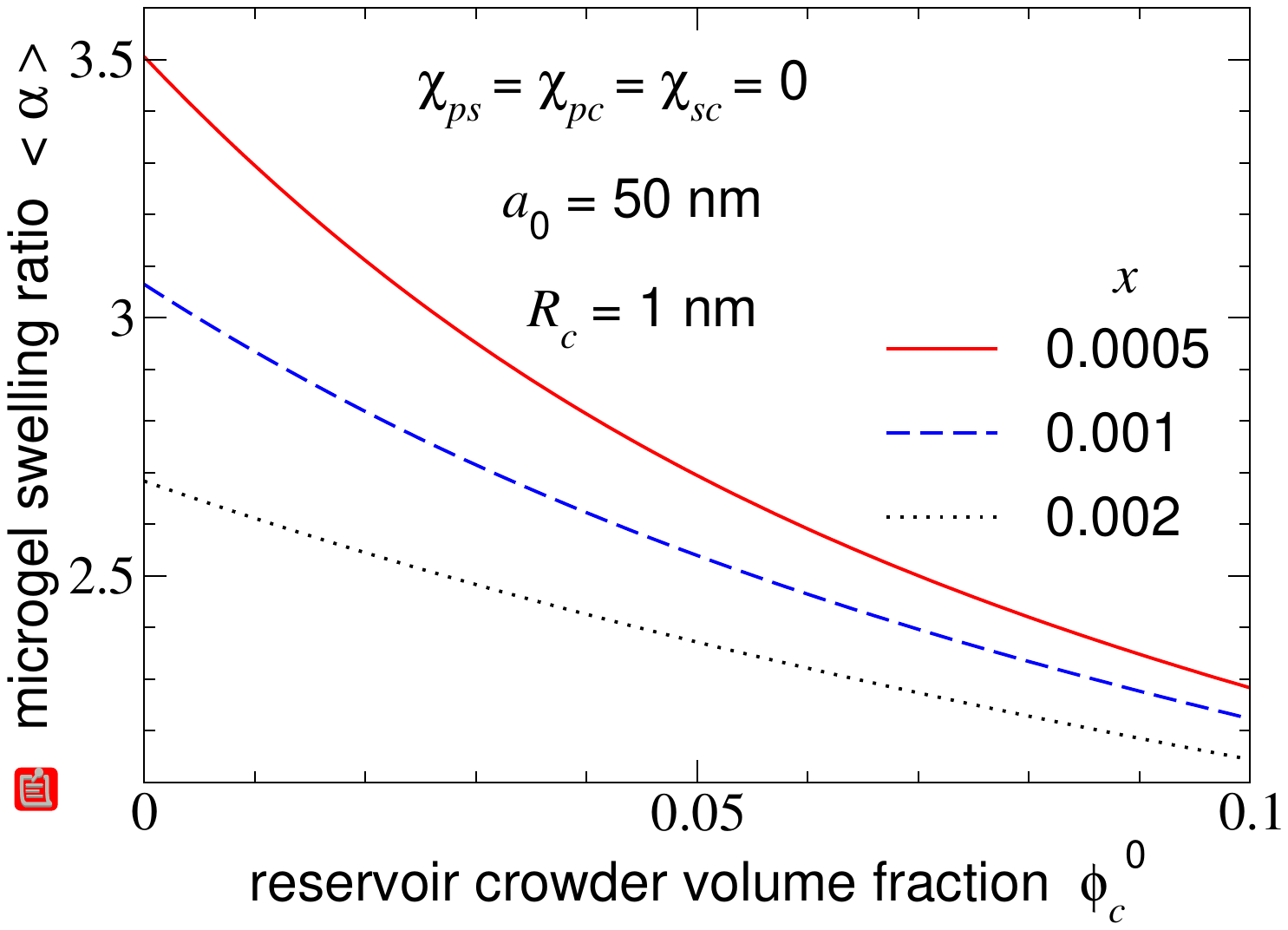}
\vspace*{-0.7cm}
\caption{Numerical results for equilibrium mean swelling ratio $\la\alpha\ra$ of a bulk hydrogel
or single microgel immersed in a solution of crowders versus reservoir volume fraction $\phi_c^0$ 
for microgel dry radius $a_0=50$ nm, crowder radius $R_c=1$ nm, and cross-link fractions 
$x=0.0005, 0.001, 0.0015$.
Flory solvency parameters are all set to zero for good solvent conditions.
}
\label{deswelling_x}
\end{figure}

\begin{figure}
\includegraphics[width=\columnwidth]{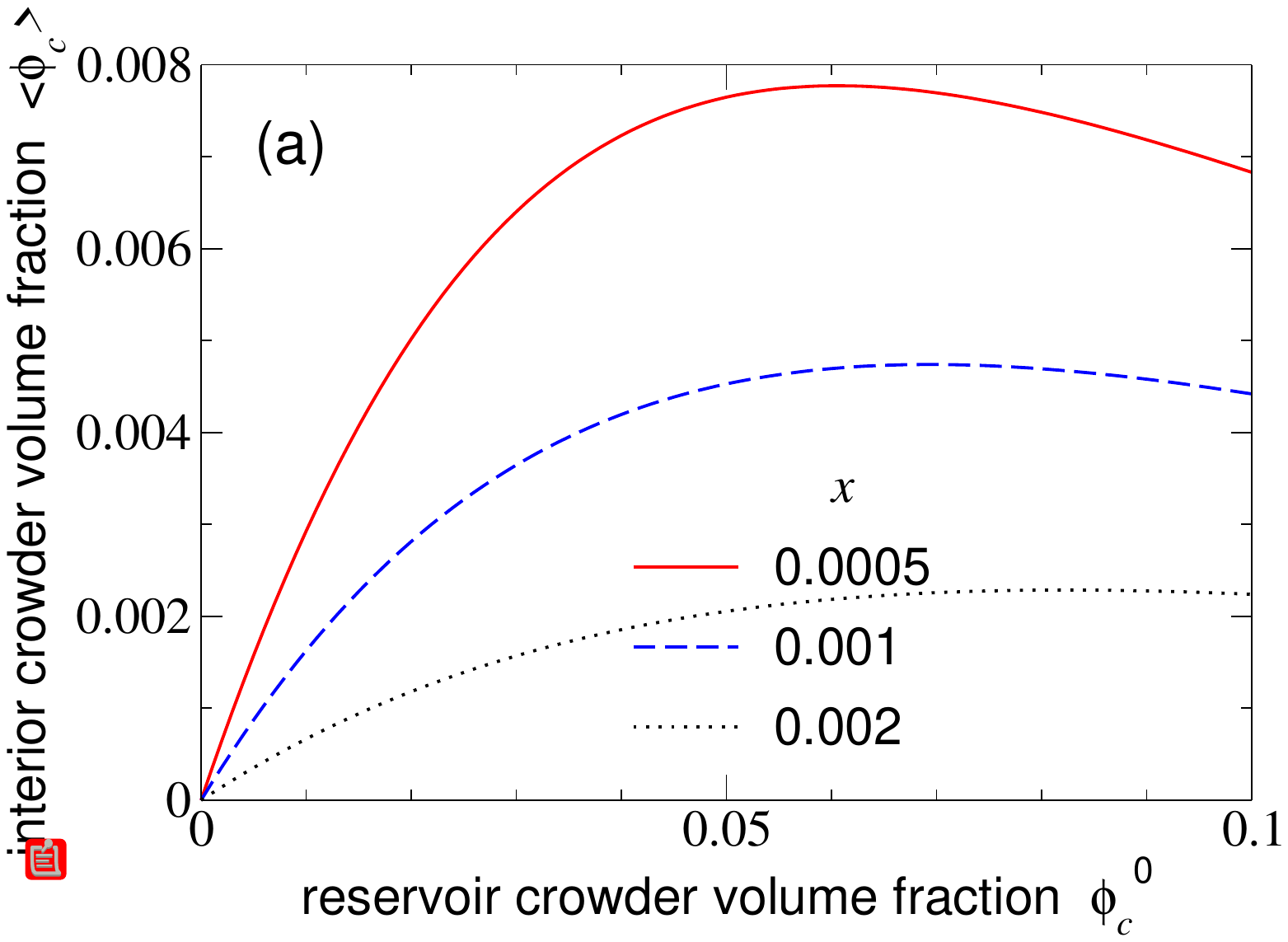}
\\[-7ex]
\includegraphics[width=\columnwidth]{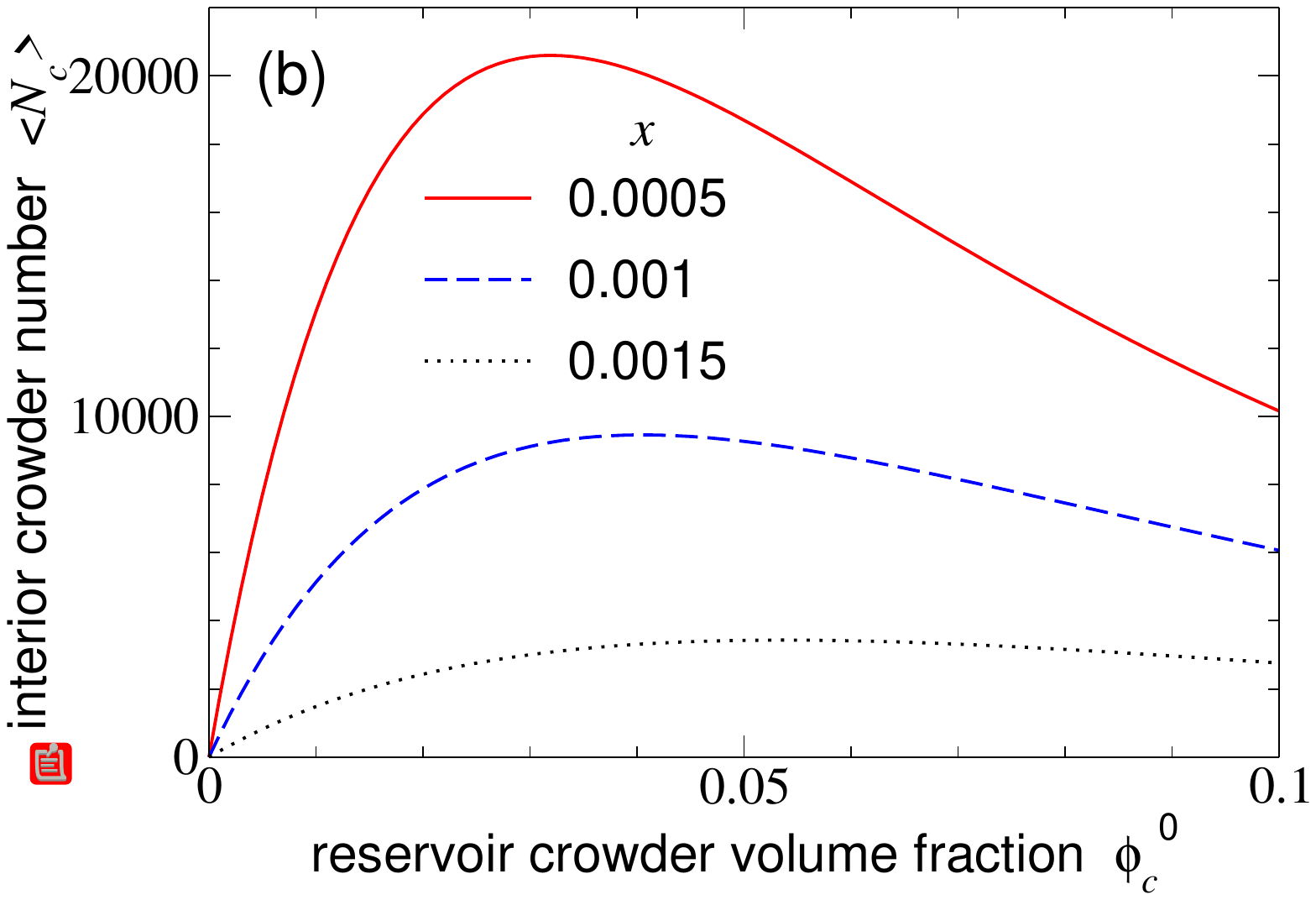}
\vspace*{-0.7cm}
\caption{Numerical results for equilibrium crowder uptake by a single microgel:
interior crowder (a) mean volume fraction $\la\phi_c\ra$ and (b) mean number $\la N_c\ra$ versus 
reservoir volume fraction $\phi_c^0$ for cross-link fractions $x=0.0005, 0.001, 0.0015$.
Other system parameters are same as in Fig.~\ref{deswelling_x}. 
}
\label{uptake_x}
\end{figure}

\begin{figure}
\includegraphics[width=\columnwidth,angle=0]{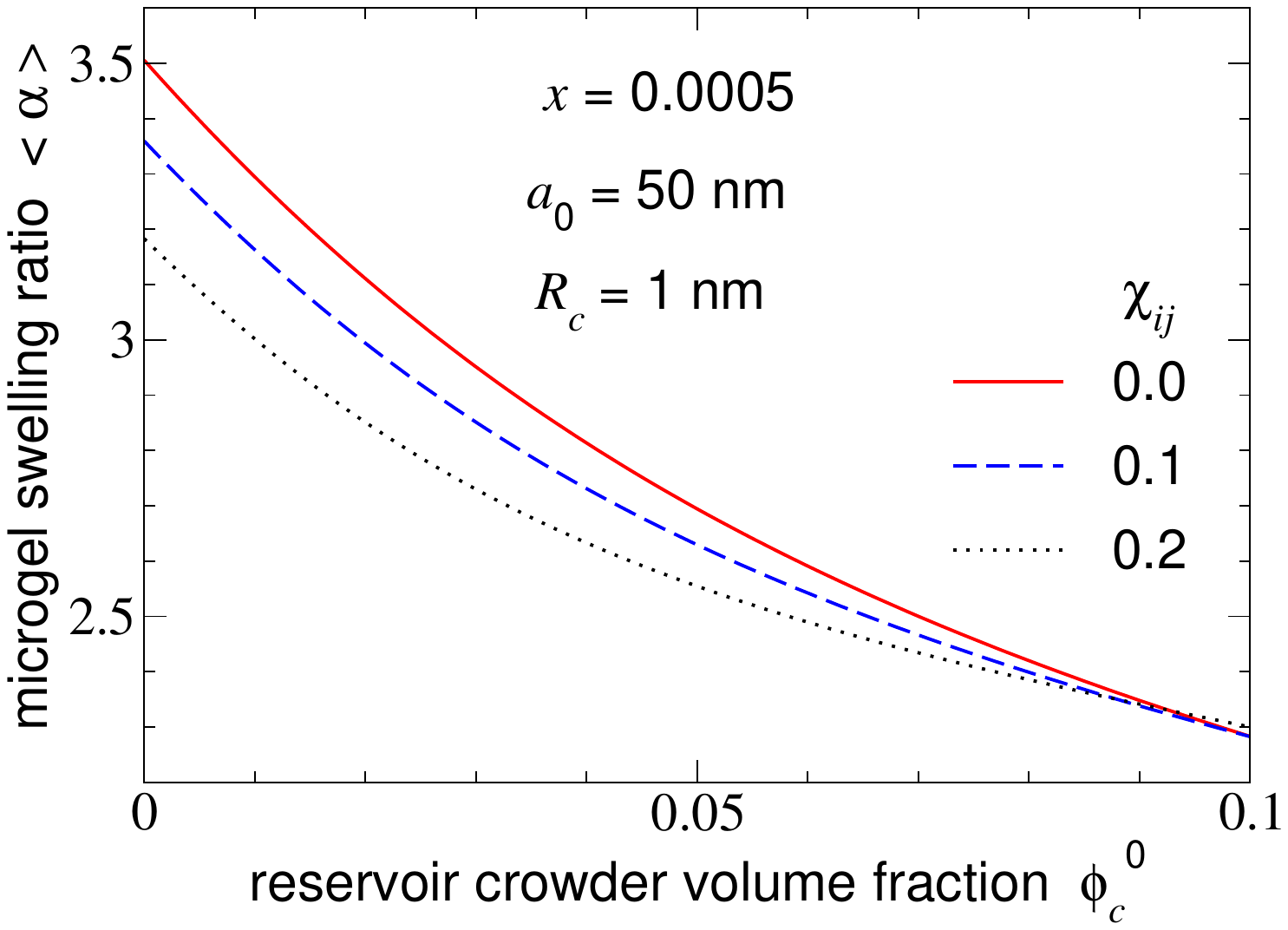}
\vspace*{-0.7cm}
\caption{Numerical results for equilibrium mean swelling ratio $\la\alpha\ra$ of a bulk hydrogel
or single microgel immersed in a solution of crowders versus reservoir volume fraction $\phi_c^0$ 
for microgel dry radius $a_0=50$ nm, cross-link fraction $x=0.0005$, and crowder radius
$R_c=1$ nm. All three solvency parameters are assumed equal ($\chi_{ps}=\chi_{pc}=\chi_{sc}$)
and set to $\chi_{ij}=0, 0.1, 0.2$.
}
\label{deswelling_chi}
\end{figure}

\begin{figure}
\includegraphics[width=\columnwidth]{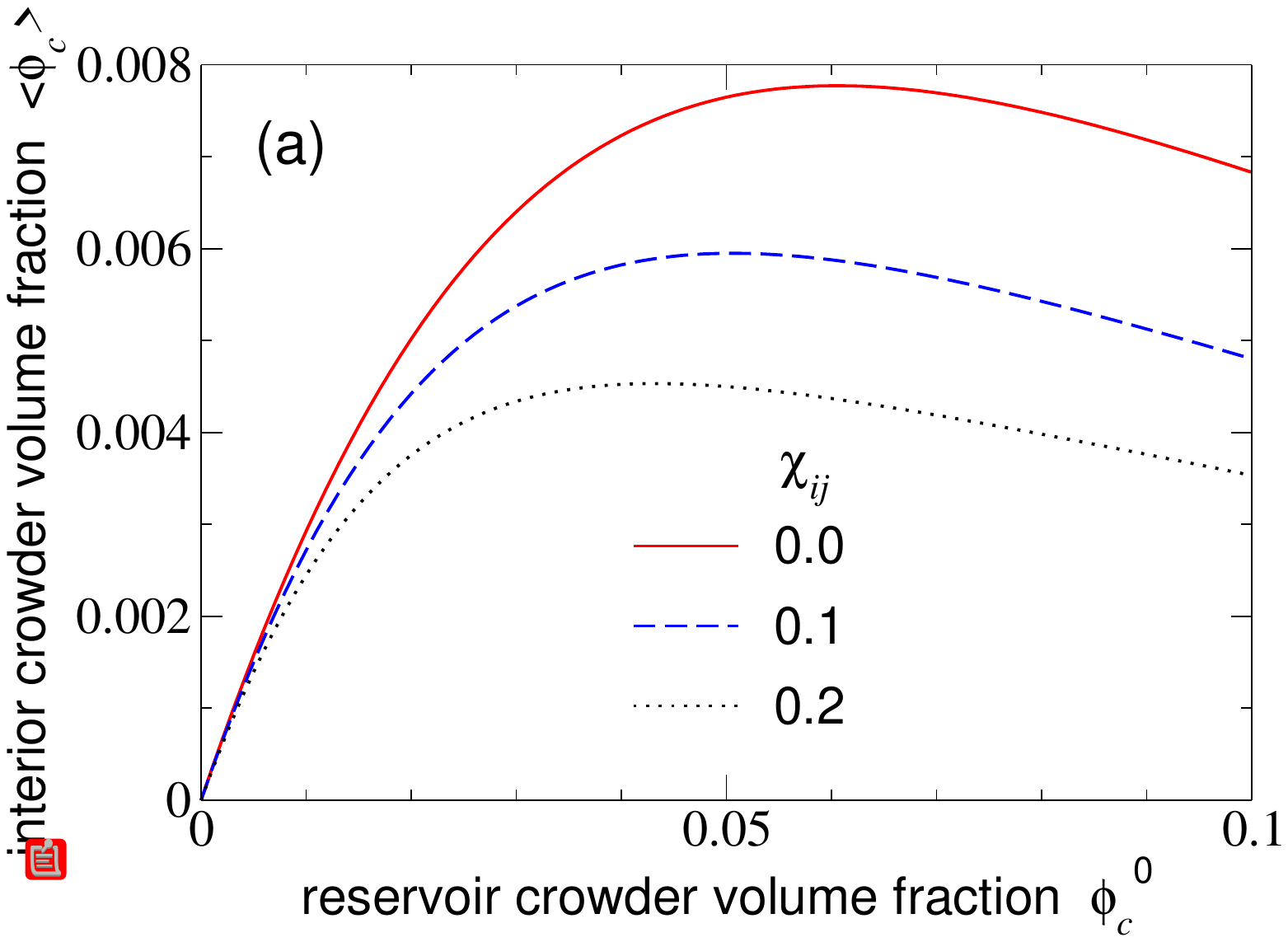}
\\[-7ex]
\includegraphics[width=\columnwidth]{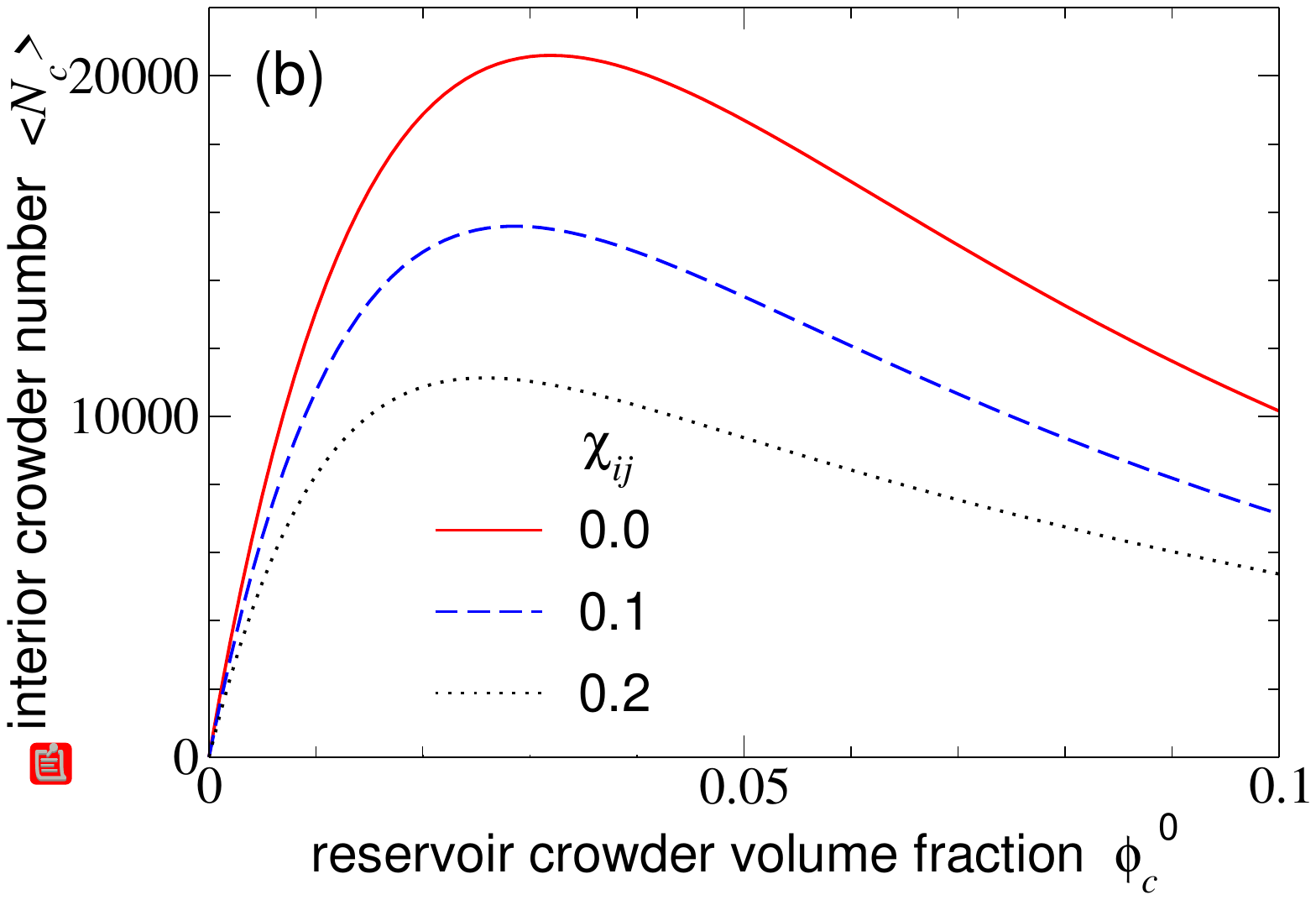}
\vspace*{-0.7cm}
\caption{Numerical results for equilibrium crowder uptake by a single microgel:
interior crowder mean (a) volume fraction $\la\phi_c\ra$ and (b) number $\la N_c\ra$ versus 
reservoir volume fraction $\phi_c^0$ for 
same system parameters as in Fig.~\ref{deswelling_chi}. 
}
\label{uptake_chi}
\end{figure}

\begin{figure}
\includegraphics[width=\columnwidth,angle=0]{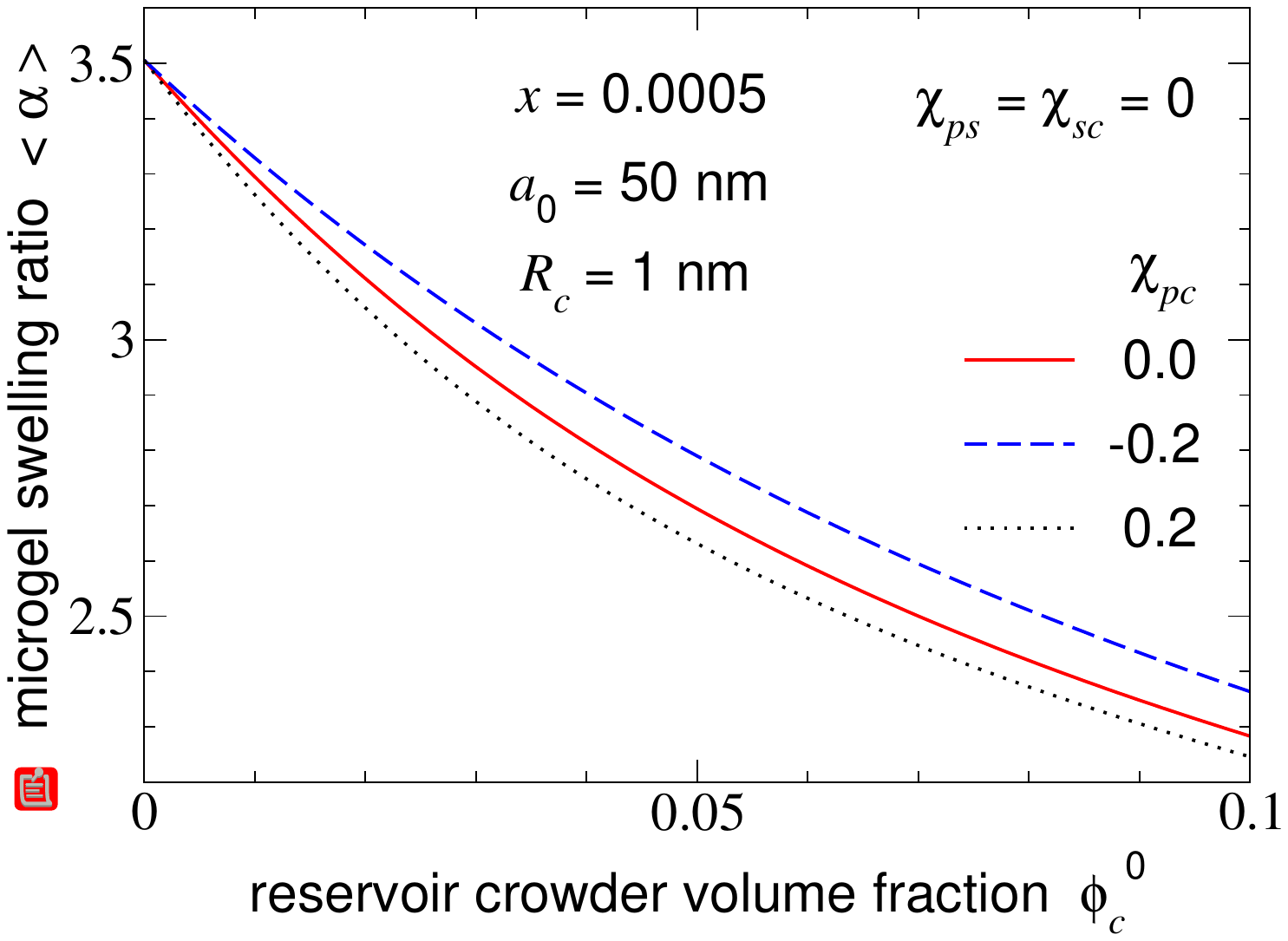}
\vspace*{-0.7cm}
\caption{Numerical results for equilibrium mean swelling ratio $\la\alpha\ra$ of a bulk hydrogel
or single microgel immersed in a solution of crowders versus reservoir volume fraction $\phi_c^0$ 
for same system parameters as in Fig.~\ref{deswelling_chi}.
Solvency parameters are set to $\chi_{ps}=\chi_{sc}=0$ and $\chi_{pc}=0, -0.2, 0.2$
to mimic effective polymer-crowder attractive and repulsive interactions.
}
\label{deswelling_chipc}
\end{figure}

\begin{figure}
\includegraphics[width=\columnwidth]{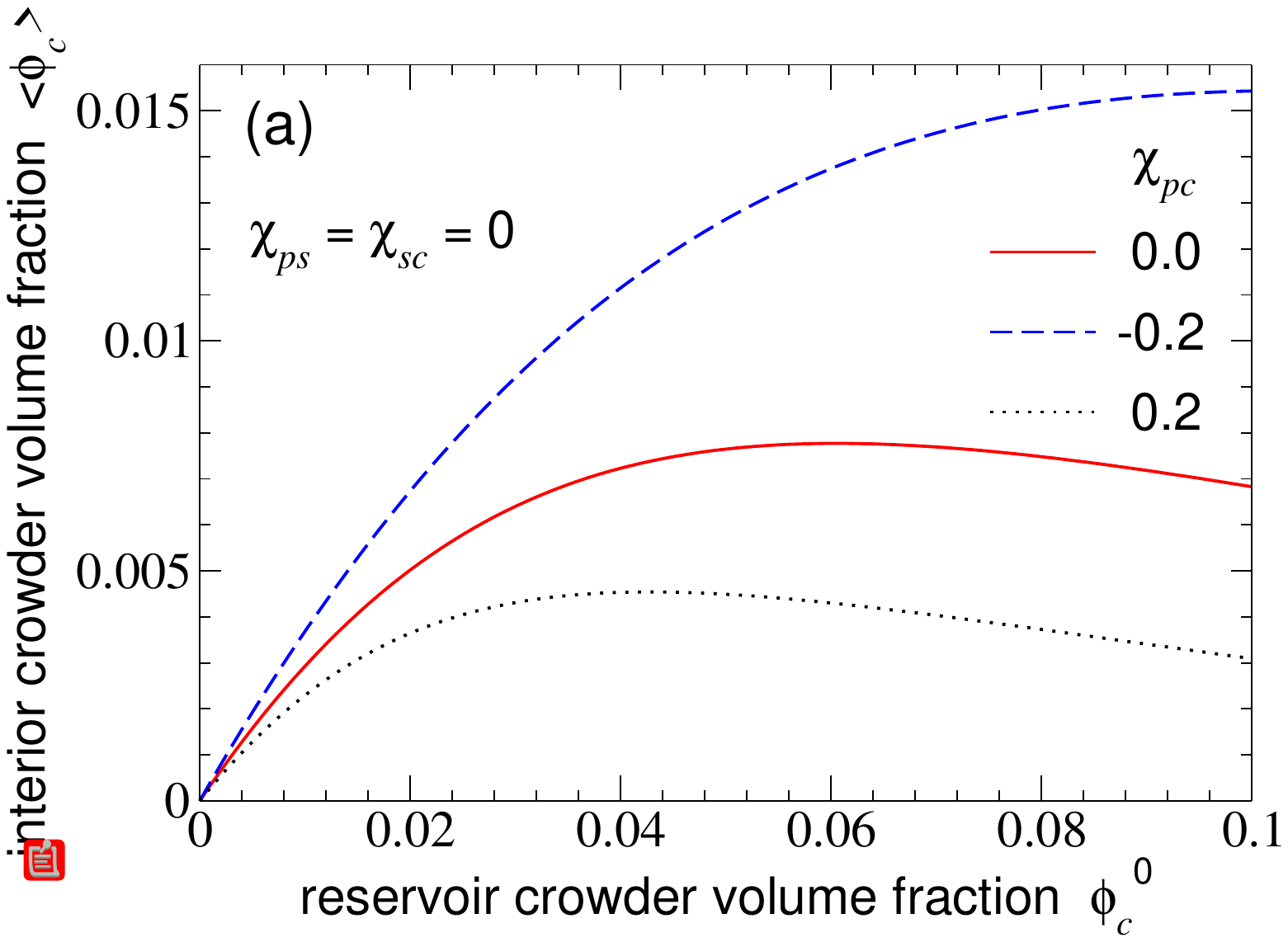}
\\[-7ex]
\includegraphics[width=\columnwidth]{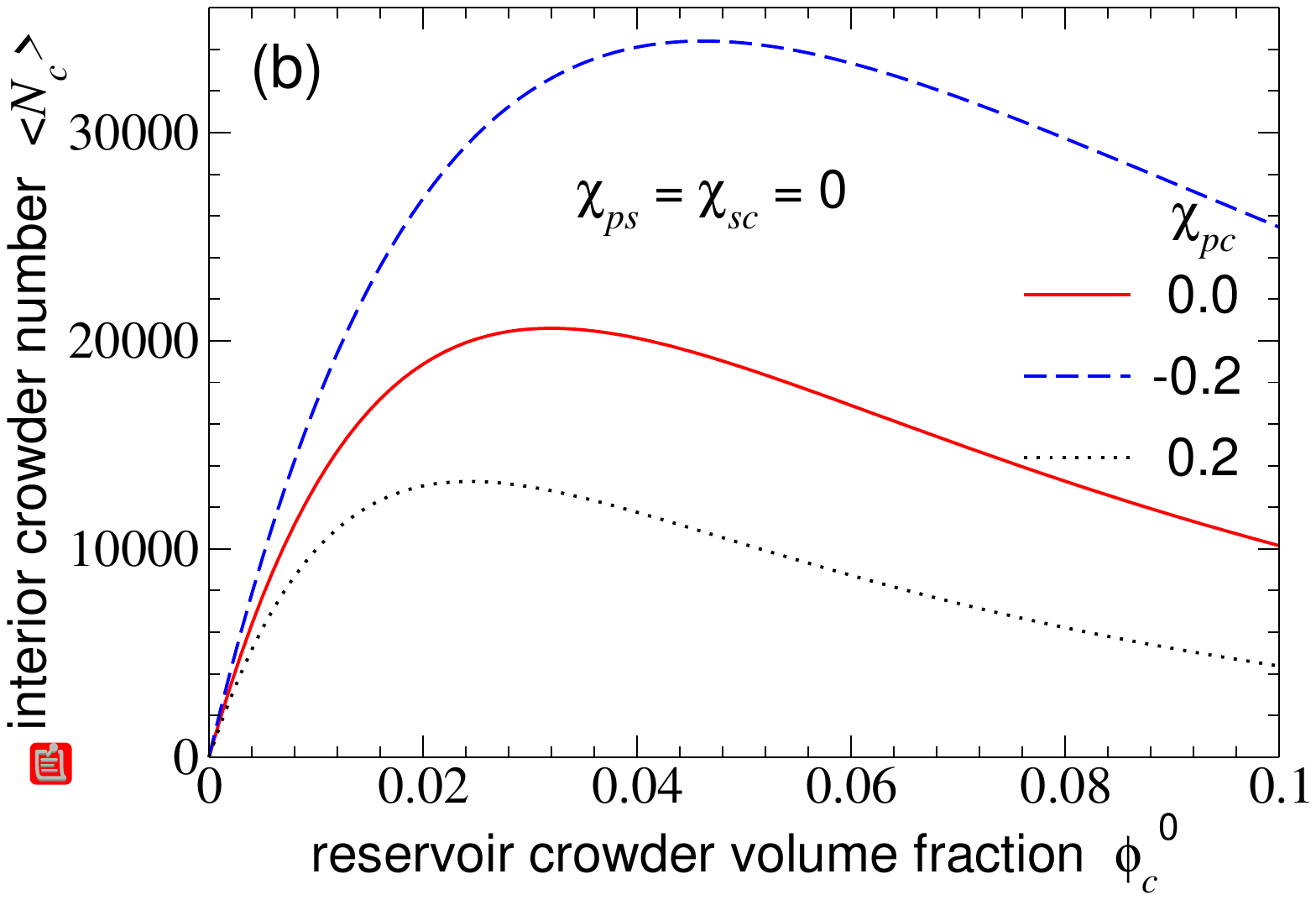}
\vspace*{-0.7cm}
\caption{Numerical results for equilibrium crowder uptake by a single microgel:
interior crowder (a) mean volume fraction $\la\phi_c\ra$ and (b) mean number $\la N_c\ra$ versus 
reservoir volume fraction $\phi_c^0$ for same system parameters as in Fig.~\ref{deswelling_chipc}.
}
\label{uptake_chipc}
\end{figure}

To illustrate the dependence of swelling and uptake on cross-link fraction and solvent quality,
we show in Figs.~\ref{deswelling_x}-\ref{uptake_chipc} our numerical results for $\la\alpha\ra$, 
$\la\phi_c\ra$, and $\la N_c\ra$ over ranges of $x$ and $\chi$. With increasing $x$, as the polymer 
network stiffens, the equilibrium mean swelling ratio decreases (Fig.~\ref{deswelling_x}), implying 
smaller mean pore size. Correspondingly, the uptake of crowders is reduced, as indicated by 
lower values of interior crowder mean volume fraction and number (Fig.~\ref{uptake_x}). 
These trends are consistent with a shrinking polymer chain $R_g$ and rising penetration free energy 
with increasing cross-link fraction. 

Upon reducing solvent quality for all species, achieved by increasing all solvency parameters 
(assuming $\chi_{ps}=\chi_{pc}=\chi_{sc}$), similar trends are observed for swelling and uptake 
as the polymer chains compactify (Fig.~\ref{deswelling_chi} and \ref{uptake_chi}). However, if 
only the polymer-crowder solvency parameter is varied [Fig.~\ref{uptake_chipc}], then
for $\chi_{pc}<0$ (mimicking effective polymer-crowder attraction) deswelling is reduced and 
uptake is enhanced, while for $\chi_{pc}>0$ (effective repulsion) the trends are reversed 
-- deswelling is enhanced and uptake is reduced. 

\begin{figure}
\includegraphics[width=\columnwidth]{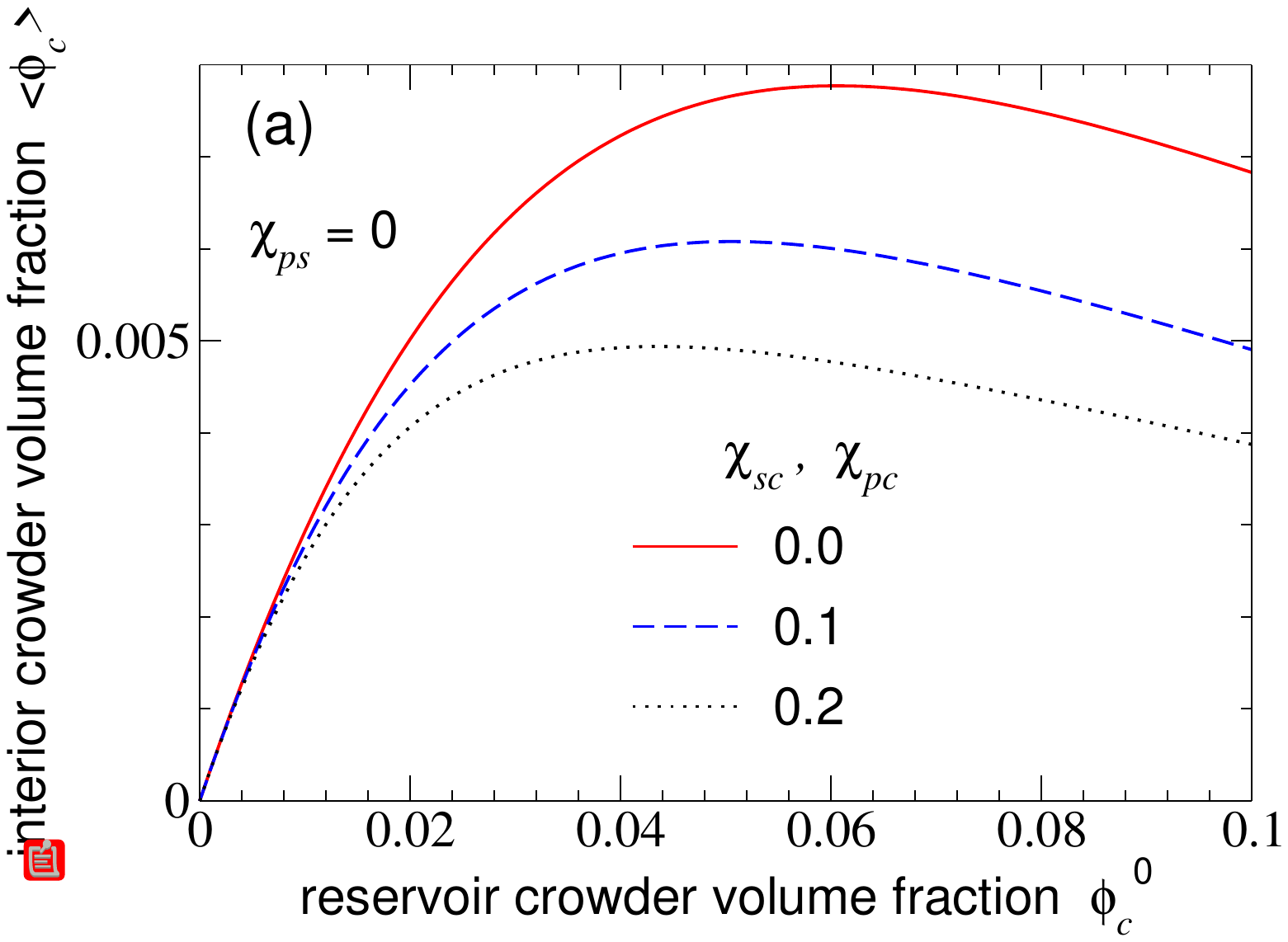}
\\[-7ex]
\includegraphics[width=\columnwidth]{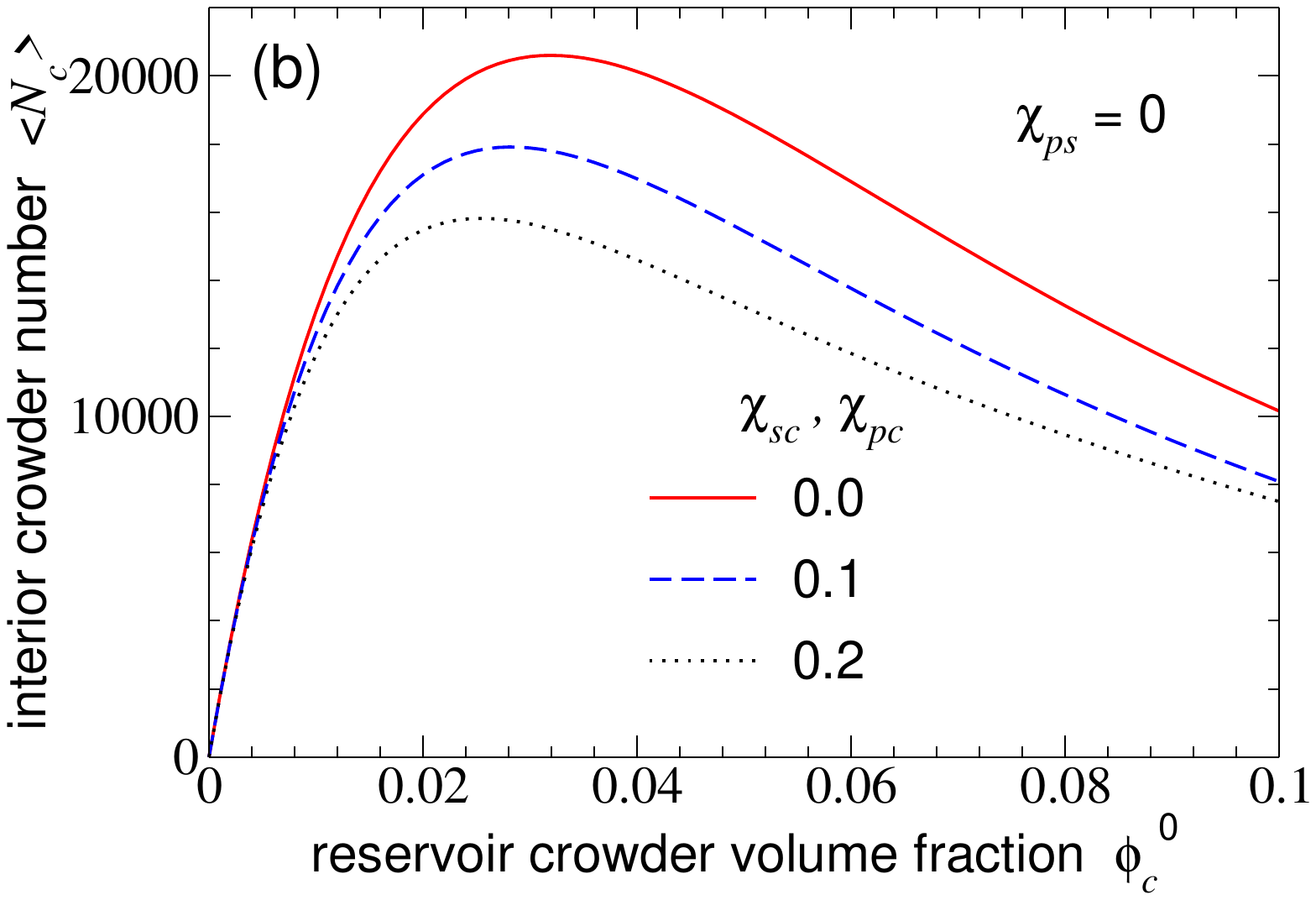}
\vspace*{-0.7cm}
\caption{Numerical results for equilibrium crowder uptake by a single microgel:
interior crowder (a) mean volume fraction $\la\phi_c\ra$ and (b) mean number $\la N_c\ra$ versus 
reservoir volume fraction $\phi_c^0$ for same system parameters as in Fig.~\ref{deswelling_chi}.
Solvency parameters are set to $\chi_{ps}=0$ and $\chi_{sc}=\chi_{pc}=0, 0.1, 0.2$ to mimic
crowder-crowder effective attraction.
}
\label{uptake_chisc_chipc}
\end{figure}

Within our model, we can also mimic mutual effective attractions between crowders
[$\epsilon_{cc}<0$ in Eq.~(\ref{Emean})], added to excluded-volume repulsion, 
by assigning positive values to both $\chi_{sc}$ and $\chi_{pc}$.
As shown in Fig.~\ref{uptake_chisc_chipc}, with increasing $\chi_{sc}$ and $\chi_{pc}$,
uptake of crowders is steadily suppressed. This predicted trend is consistent with the
energetic cost incurred by crowders mixing with polymer inside the microgel instead of with 
one another outside.

\begin{figure}
\includegraphics[width=\columnwidth,angle=0]{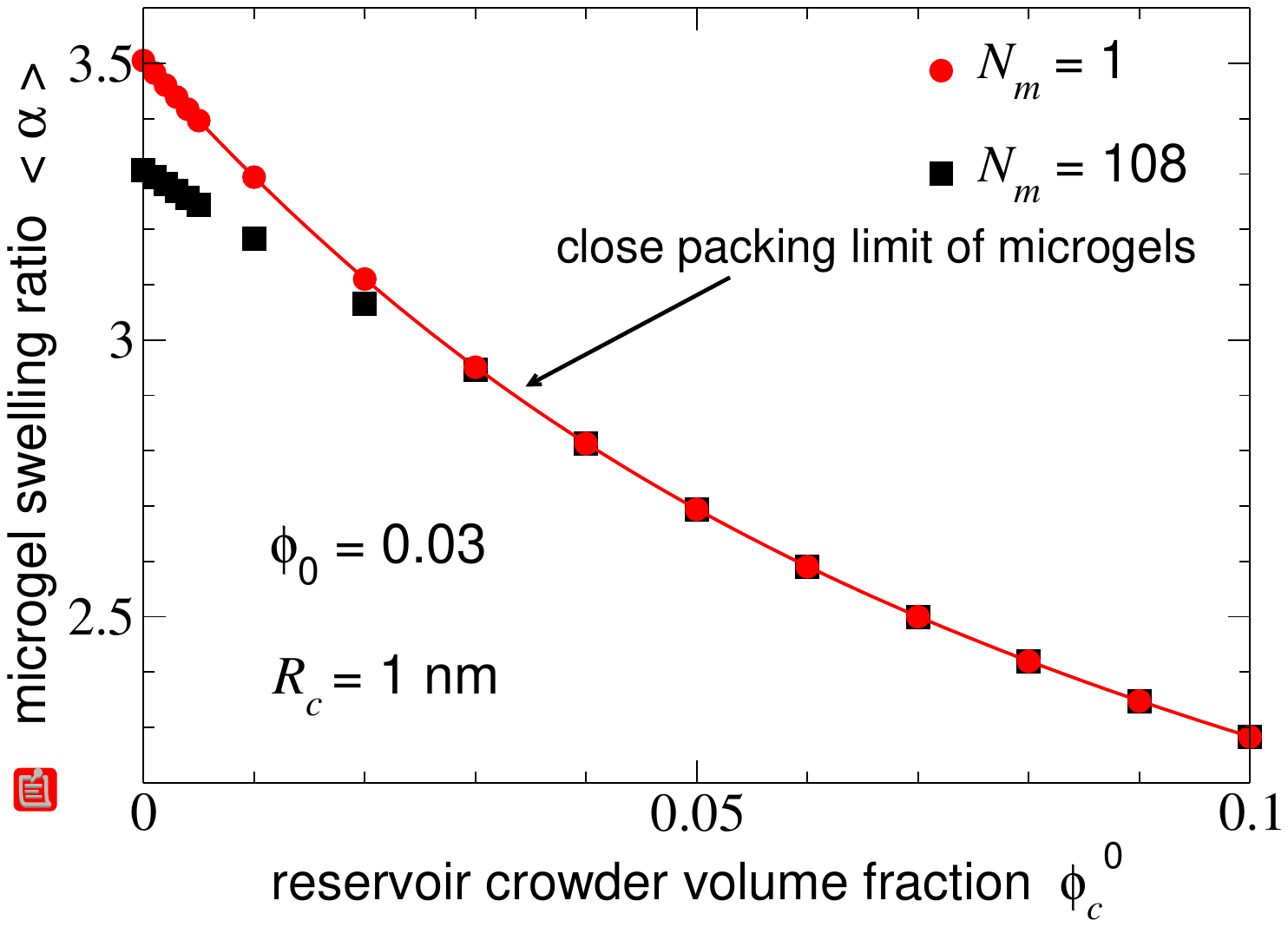}
\vspace*{-0.7cm}
\caption{MC simulation data for equilibrium mean swelling ratio $\la\alpha\ra$ of a single 
isolated microgel (red circles) and of $N_m=108$ microgels (black squares) dispersed in 
a solution of crowders of radius $R_c=1$ nm versus reservoir volume fraction $\phi_c^0$. 
The dry volume fraction of microgels in the suspension is $\phi_0=0.03$. 
Numerical results are also shown for $N_m=1$ (curve). 
Other system parameters are same as in Fig.~\ref{deswelling_Rc}. 
}
\label{deswelling_N108}
\end{figure}

\begin{figure}
\includegraphics[width=\columnwidth]{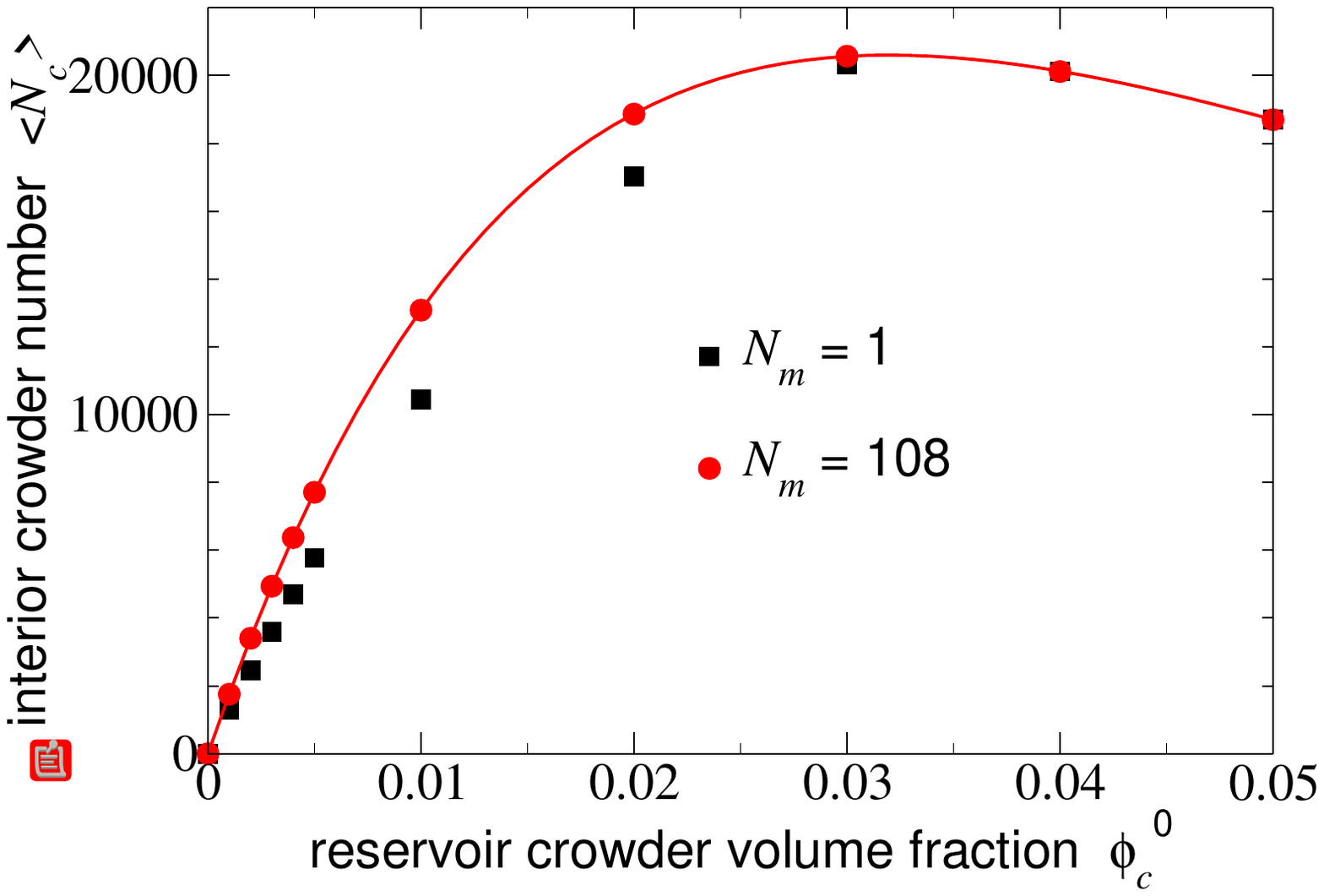}
\vspace*{-0.7cm}
\caption{MC simulation data for equilibrium mean number of crowders $\la N_c\ra$ inside an 
isolated microgel (red circles) and for $N_m=108$ microgels (black squares) dispersed in a 
solution of crowders of radius $R_c=1$ nm versus reservoir volume fraction $\phi_c^0$.
Numerical results are also shown for $N_m=1$ (curve). 
Other system parameters are same as in Fig.~\ref{deswelling_N108}. 
}
\label{uptake_N108}
\end{figure}

To investigate the influence of interparticle interactions on single-microgel and bulk structural 
properties in a suspension, we performed simulations of $N_m=108$ interacting microgels dispersed 
in a solution of crowders. In the Hertz potential [Eqs.~(\ref{vH})-(\ref{Bij2})], we set the 
calibration factor to $Y_c=10$, yielding a Young's modulus of $Y\simeq 50$ kPa for relatively stiff
microgels~\cite{nieves-bulk-shear-pre2011}. 
From our simulations, we averaged over microgels and MC steps to compute equilibrium values of the 
mean swelling ratio and interior number of crowders. Figures~\ref{deswelling_N108} and 
\ref{uptake_N108} compare our data for an isolated microgel with corresponding data for a suspension.
At lower reservoir concentrations ($\phi_c^0<0.04$), where for these system parameters the particles 
are so swollen that they remain in an FCC crystal configuration, Hertzian interactions between 
neighbors evidently contribute additional deswelling, especially when the microgel density exceeds 
that of close-packed spheres. Concomitantly, interparticle interactions reduce uptake of crowders. 
At higher reservoir concentrations ($\phi_c^0>0.05$), the microgels deswell to the point that 
interparticle correlations become sufficiently weak as to destabilize (melt) the crystal. In this range, 
microgels in a suspension have essentially the same swelling ratio and crowder uptake as an 
isolated microgel, whose deswelling is purely crowder-induced. 

\begin{figure}
\includegraphics[width=\columnwidth,angle=0]{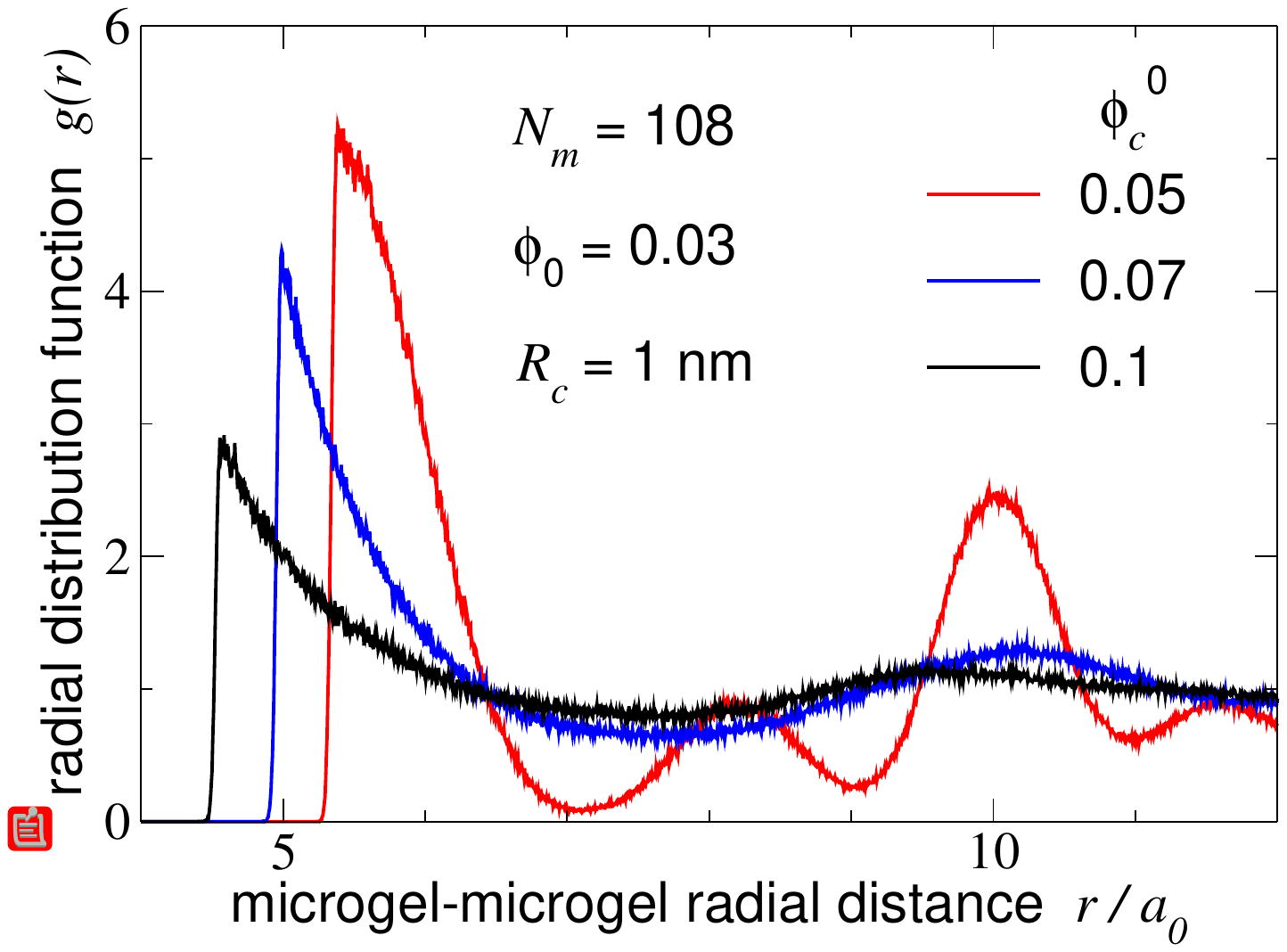}
\vspace*{-0.7cm}
\caption{MC simulation data for radial distribution function $g(r)$ vs. radial distance $r$ 
(units of microgel dry radius) for $N_m=108$ microgels dispersed in crowder solutions
of reservoir concentrations $\phi_c^0=0.05$ (red), 0.07 (blue), and 0.1 (black). The volume
fraction of the microgels in their dry state is $\phi_0=0.03$. Other system parameters 
are same as in Fig.~\ref{deswelling_Rc}. With increasing $\phi_c^0$, the particles deswell,
as reflected by the shift of the main peak toward shorter distances. As a result, pair correlations 
weaken, indicated by lower and broader peaks, and the suspension transitions from a solid 
(FCC crystal) to a fluid phase.
}
\label{rdf}
\end{figure}

\begin{figure}
\includegraphics[width=\columnwidth,angle=0]{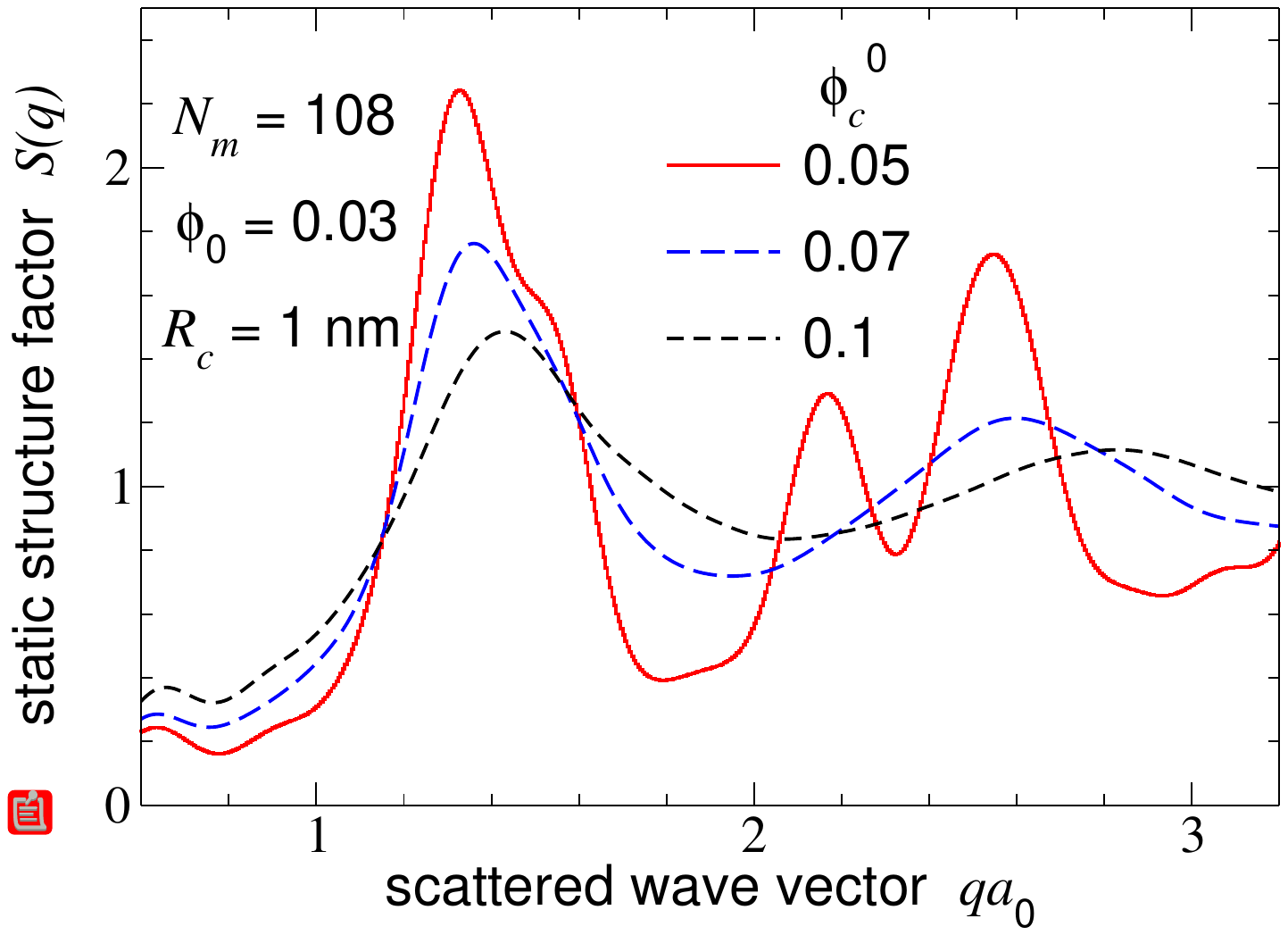}
\vspace*{-0.7cm}
\caption{MC simulation data for static structure factor $S(q)$ as a function of scattered wave 
vector $q$ (units of inverse microgel dry radius) for $N_m=108$ microgels dispersed in 
crowder solutions of reservoir concentrations $\phi_c^0=0.05$ (solid red curve), 0.07 (long-dashed blue), 
and 0.1 (short-dashed black). The volume fraction of the microgels in their dry state is $\phi_0=0.03$. 
Other system parameters are same as in Fig.~\ref{deswelling_Rc}. Consistent with the results for $g(r)$
shown in Fig.~\ref{rdf}, with increasing $\phi_c^0$, pair correlations become weaker and the 
solid phase is destabilized. Thus, crowder uptake weakens correlations and induces melting.
}
\label{ssf}
\end{figure}

From our simulations, we further computed the radial distribution function $g(r)$ by histogramming the 
center-to-center separation $r$ between pairs of particles and averaging over configurations. 
A measure of pair correlations, $g(r)$ is proportional to the conditional probability of finding 
a particle at radial distance $r$ away from another particle~\cite{HansenMcDonald}.
We also computed the orientationally averaged static structure factor, 
\begin{equation}
S(q)=1+\frac{2}{N_m}\sum_{i<j=1}^{N_m}\la\frac{\sin(qr_{ij})}{r_{ij}}\ra,
\label{Sq}
\end{equation}
which is proportional to the intensity of scattered light, $q$ being the magnitude of the 
scattered wave vector, and to the Fourier transform of $g(r)$.

The phenomenon of crowding-induced melting is confirmed by comparing radial distribution functions
and static structure factors of suspensions at different reservoir concentrations. As shown in
Figs.~\ref{rdf} and \ref{ssf}, at $\phi_c^0=0.05$, $g(r)$ and $S(q)$ exhibit significant structure, 
reflecting strong pair correlations indicative of long-range periodic ordering in a stable or
metastable crystalline phase. At higher $\phi_c^0$, these correlations weaken as the crystal 
gives way to the fluid phase. We emphasize that the melting transition observed here, at constant 
temperature and microgel density, is driven entirely by crowding-induced deswelling,  
and not by any crowder-microgel correlations, which are absent from our coarse-grained model.

\section{Conclusions}\label{conclusions}
By developing and implementing a theory of cross-linked polymer networks immersed in a solution 
of macromolecular crowding agents, we modeled the response of hydrogels and microgels to volume-excluding 
crowders. From a free energy composed of mixing, elastic, and penetration entropy contributions, 
as well as mixing energy in a mean field approximation, we derived the chemical potentials of solvent 
and crowders. By equating chemical potentials of both species inside and outside the gel, we computed 
the equilibrium swelling ratio and crowder uptake concentration over a range of reservoir solution
concentration, crowder size, cross-link fraction, and solvency parameters. 

To validate our numerical methods, we also performed Monte Carlo simulations with trial changes 
in swelling ratio and interior crowder concentration. The simulations effectively minimize the
Landau free energy of the system, which is thermodynamically equivalent to equating 
chemical potentials in the interior and exterior phases. By adding a Hertzian elastic pair potential 
to our model, we simulated suspensions of microgels and computed the influence of interparticle
interactions on swelling and uptake and of crowding on microgel pair correlations.

Our results demonstrate how macromolecular crowders that exclude volume, when absorbed from solution
by polymeric gels, can induce deswelling, ever more strongly with decreasing crowder size. 
Our analysis reveals that mixing entropy favors uptake of crowders at low reservoir concentrations, 
but that with increasing concentration, uptake first peaks and then declines. The peak value of 
the uptake concentration decreases with increasing crowder size. We further conclude that hard crowders, 
by driving deswelling, weaken pair correlations in microgel suspensions and can induce 
melting of a crystal.
  
Finally, we summarize the scope and limitations of our theoretical framework. First, our coarse-grained 
model of polymeric gels neglects microscopic details in favor of capturing the dominant physics. 
Second, our theory of microgel-crowder solutions is based on a lattice model and a Flory-Huggins 
mean-field approximation for the mixing free energy. While our theory approximately accounts for 
crowder-crowder pair correlations, it neglects many-body correlations and so applies only to low 
crowder concentrations. Third, we model crowders as idealized hard spheres that exclude volume 
and represent them by a continuous density field rather than as discrete particles. Fourth, we describe 
swelling of gels by the Flory-Rehner mean-field theory, which assumes affine swelling 
and Gaussian polymer chains, and interactions between microgels by a simple elastic Hertzian potential. 
Although we focus here on uniformly cross-linked microgels, our theory can be naturally extended to 
heterogeneous microgels, for example, by means of multi-Hertzian interaction 
models~\cite{zaccarelli-multiHertzian2018,moncho-jorda2025}. Lastly, our use of polymer field theory 
to account for the free energy cost of crowder penetration neglects molecular details of polymers.

Our computationally efficient modeling approach captures the key physical features of these complex 
materials and our results highlight and demonstrate the important role of excluded-volume effects 
in macromolecular crowding of polymeric gels. The accuracy of our approximations and predictions may be 
tested via simulations of molecular-scale models~\cite{zaccarelli2017,zaccarelli2024,zaccarelli2025} 
and by experiments that add crowding agents to gel solutions and measure swelling and uptake response. 
A natural extension of our model will be to incorporate flexible crowders with freedom to change 
conformation in response to confinement within the pores of a gel. Future work may also include adapting 
the modeling to describe ionic microgels~\cite{levin2002,denton-cecam2014,colla-likos2014,moncho-jorda2013,
quesada-perez-moncho-jorda-anta2015,moncho-jorda-dzubiella2016} and study the interplay between 
crowding-induced and charge-induced swelling. 

\acknowledgments
\noindent This work was supported by the National Science Foundation under Grant No.~DMR-1928073.

\newpage
\begin{center}
{\bf AUTHOR DECLARATION} 
\end{center}

\noindent {\bf Conflict of Interest} \\[1ex]
The authors have no conflicts to disclose.
\\[3ex]
{\bf Author Contributions} \\[2ex]
{\bf Mahesh Aryal:}
formal analysis (equal); methodology (equal); calculations (equal); software (equal); 
data curation (equal); visualization (equal); validation (equal); writing -- original draft (equal); 
writing -- review and editing (supporting). 
\\[1ex]
{\bf Alan R.~Denton:}
conceptualization (lead); project administration (lead); resources (lead); supervision (lead); 
formal analysis (equal); methodology (equal); calculations (equal); software (equal); 
data curation (equal); visualization (equal); validation (equal); writing -- original draft (equal); 
writing -- review and editing (lead). 

\vspace*{0.3cm}
\begin{center}
{\bf DATA AVAILABILITY} 
\end{center}

\noindent Data supporting the findings of this study are available from the corresponding author
upon reasonable request.


\bibliography{microgels-resub}

\end{document}